\documentclass[aps,prd,twocolumn,prd,showpacs,showkeys,preprintnumbers,bibnotes,floatfix,longbibliography,notitlepage,nofootinbib,superscriptaddress,superscriptgaddress,
]{revtex4-2}

\pdfoutput=1
\usepackage{amsmath}
\usepackage{amsfonts}
\usepackage{amssymb}
\usepackage{mathrsfs}
\usepackage{graphicx}
\usepackage{color}
\usepackage{bbding}
\usepackage{tabularx}
\usepackage[usenames,dvipsnames,table]{xcolor}
\usepackage[normalem]{ulem}
\usepackage{makecell}
\usepackage{slashed}
\usepackage{soul}

\usepackage[hidelinks]{hyperref}
\usepackage{multirow}
\usepackage{makecell}
\usepackage{upgreek}
\usepackage[capitalise]{cleveref}
\usepackage{subfigure}
\usepackage[normalem]{ulem}
\begin{document}

\title{Supernova cooling from neutrino-devouring dark matter}
\author{Yugen Lin}
\email{linyugen@itp.ac.cn}
\affiliation{Institute of Theoretical Physics, Chinese Academy of Sciences, Beijing, 100190, China}

\author{Chih-Ting Lu}
\email{ctlu@njnu.edu.cn}
\affiliation{Department of Physics and Institute of Theoretical Physics, Nanjing Normal University,
Nanjing, 210023, China}
\affiliation{Nanjing Key Laboratory of Particle Physics and Astrophysics, Nanjing, 210023, China}

\author{Ningqiang Song}
\email{songnq@itp.ac.cn}
\affiliation{Institute of Theoretical Physics, Chinese Academy of Sciences, Beijing, 100190, China}

%\date{\today}

\begin{abstract}
Supernova cooling provides a powerful probe of physics beyond the Standard Model (SM), in particular for new, light states interacting feebly with SM particles. In this \textit{Letter}, we investigate for the first time the production of fermionic dark matter (DM) via the neutrino-devouring process inside a core-collapse supernova, which contributes to the excessive cooling of the supernova. By incorporating state-of-the-art supernova simulation data and the full time evolution information, we derive stringent and robust limits on DM interactions, which can improve upon direct detection limits by several orders of magnitude. We exclude the cross sections down to $10^{-51}-10^{-58}$~cm$^2$ in the keV-MeV mass range for DM-electron scattering, and $10^{-49}-10^{-56}$~cm$^2$ in the 0.1-100~MeV mass range for DM-nucleon scattering, supplemented by complementary constraints from cosmology, astrophysics, colliders and direct detection experiments in the larger cross section regime. We also close almost the entire window in which fermionic DM constitutes $\mathcal{O}(1)$ fraction of DM due to its coupling to electrons in the keV-MeV mass range.
\end{abstract}

\maketitle
%\tableofcontents

\textbf{\textit{Introduction}} ---
Astronomical and cosmological observations have provided compelling evidence for dark matter (DM), while the fundamental properties of DM remain mysterious. Especially, DM on the GeV-TeV mass scale has been subject to extensive searches in multi-tonne scale experiments underground by investigating the recoil signals of nuclei and electrons down to the energy of $\mathcal{O}$(keV) range. The DM-nucleon scattering cross section has been excluded with unprecedented precision approaching the neutrino floor in this mass range~\cite{XENON:2023cxc,LZ:2024zvo,PandaX:2024qfu}.

The null results of direct detection experiments have sparked the search for lighter DM below the GeV mass scale. However, such DM typically imparts a recoil energy that falls outside the reach of large-scale experiments, causing a sensitivity loss in the low mass end. Recently, sub-GeV fermionic DM with conversion to neutrino $\nu$ (FDMCN)~\cite{Brdar:2018qqj,Dror:2019onn,Dror:2019dib,Dror:2020czw} has attracted significant attention due to its distinct detection signatures~\cite{Brdar:2018qqj,Dror:2019onn,Dror:2019dib,Dror:2020czw,Chang:2020jwl,Hurtado:2020vlj,Li:2020pfy,Chen:2021uuw,Chao:2021bvq,Li:2022kca,Berger:2022cab,Haselschwardt:2023thp,Candela:2023rvt,Cox:2023cjw,Ma:2024aoc,Ma:2024tkt,Ge:2024euk,Ma:2024gqj,Ge:2024lzy,Ma:2025gzn,Geng:2025mtp}. Such DM of mass $M_\chi$ scatters with target $T$ through the process $\chi+T\rightarrow \nu + T$, and deposits energy $E_R\simeq M_\chi^2/(2M_T)$ in the target, leaving a peak-like recoil spectrum. The recoil energy is enhanced by a factor of $1/v_\chi^2$ and is $\mathcal{O}(10^5-10^6)$ larger than the traditional elastic scattering, which overcomes the threshold problem for light DM. This feature has motivated the experimental searches at PandaX-4T~\cite{PandaX:2022osq,PandaX:2022ood,PandaX:2024cic}, the Majorana Demonstrator~\cite{Majorana:2022gtu},  CDEX-10~\cite{CDEX:2022rxz,CDEX:2024bum}, PICO-60~\cite{PICO:2025rku} and EXO-200~\cite{EXO-200:2022adi,Richardson:2025quk}. The non-relativistic DM scattering cross section depends on the mass of DM and the target, as well as the type of interactions~\cite{Ge:2022ius}. In the limit $M_\chi\ll M_T$, it reduces to a cross section $\bar{\sigma}\equiv \sigma v_\chi \simeq M_\chi^2/4\pi\Lambda^4$, which is reported by the experimental groups.

On the other hand, core-collapse supernova provides an ideal environment for the test of new physics beyond the Standard Model (BSM). The hot and dense core of supernova facilitates the efficient production of exotic particles up to $\mathcal{O}(100)$~MeV mass, even if the particle is only weakly coupled. These particles may escape from the supernova, introducing an additional cooling channel for the protoneutron star and altering the emissivity of neutrinos~\cite{Raffelt:2006cw}. Since the detection of the neutrino burst from Supernova SN1987A~\cite{Kamiokande-II:1987idp}, supernova cooling has been employed to probe various BSM models (see e.g.~\cite{Ellis:1987pk,Raffelt:1987yt,Raffelt:1996wa,Chang:2018rso,Fiorillo:2022cdq,Li:2025beu,Cappiello:2025tws,Caputo:2025aac,Fiorillo:2023ytr}.)

\begin{figure}[!b]
    \centering    \includegraphics[width=\columnwidth]{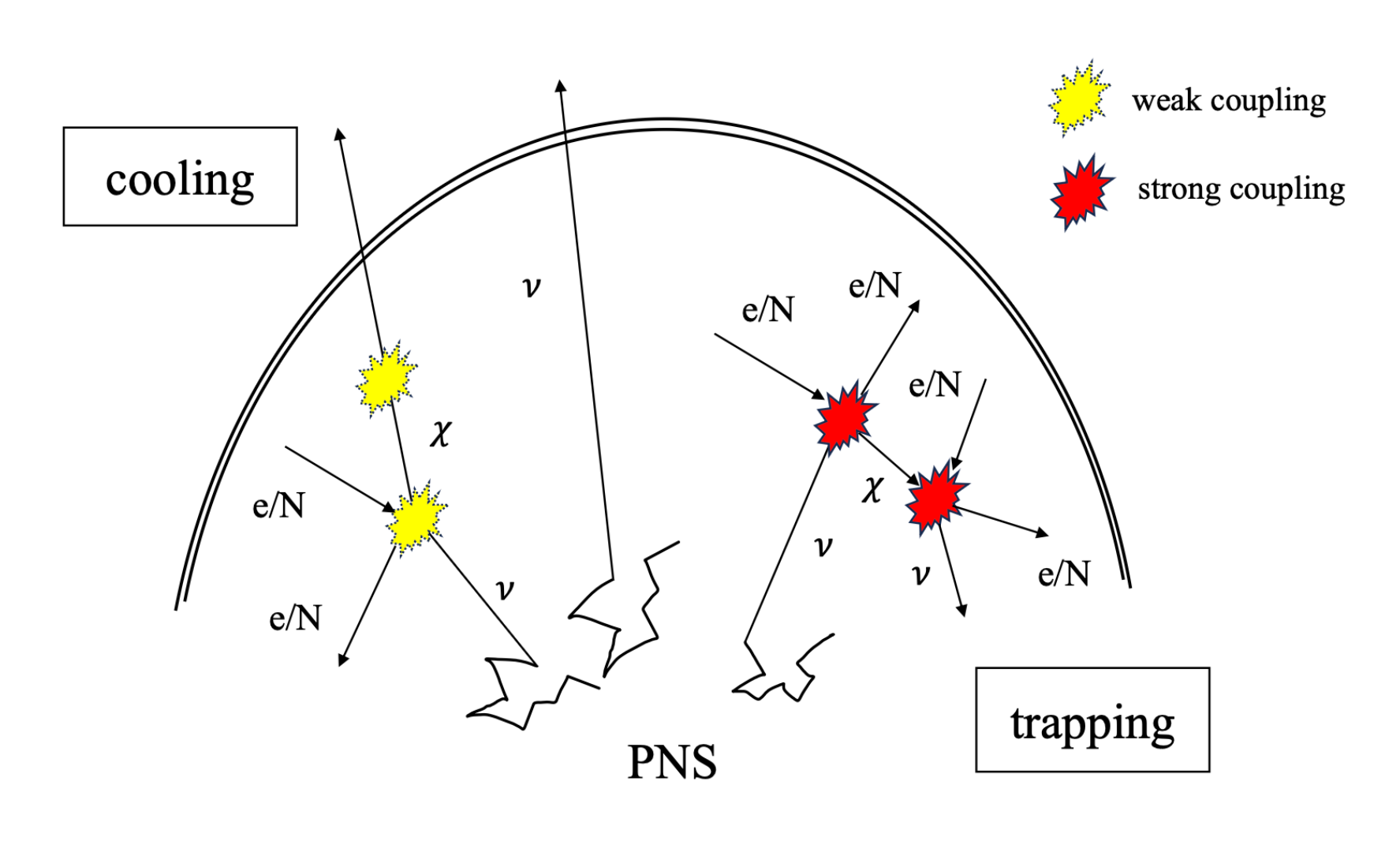}
    \caption{Illustration of the relevant processes in this work. Fermionic DM is produced when supernova neutrinos scatter with surrounding matter, causing cooling of a progenitor star in addition to neutrinos. If the coupling is too strong, the produced DM subsequently scatters and back-converts to neutrinos.}
    \label{fig:illustration}
\end{figure}

In this \textit{Letter}, for the first time, we use supernova to set stringent constraints on FDMCN. Unlike previous works looking for the production of new particles through annihilation or bremsstrahlung~\cite{Chang:2018rso,Fiorillo:2022cdq,Li:2025beu,Cappiello:2025tws}, we rather focus on the neutrino-devouring process where the supernova neutrinos scatter with the electron and nucleon to produce a fermionic DM, as illustrated in Fig.~\ref{fig:illustration}. This is the reverse process of the direct detection but significantly broadens the DM mass range without being limited by the experimental threshold or the region of interest. We have incorporated the full time evolution information of supernova from state-of-the-art simulation data in DM production and propagation to set robust constraints using the cooling criterion.

We find supernova cooling sets limits on the DM scattering cross section spanning about seven orders of magnitude. For DM-electron interaction, nearly all parameter space for FDMCN to constitute $\mathcal{O}(1)$ fraction of the cosmological DM is excluded in the keV-MeV mass window. The cooling constraints also show strong complementarity with constraints from cosmology, astrophysics, collider and direct detection experiments. For DM coupling to nucleons, supernova cooling rules out the cross section down to $10^{-56}$~cm$^2$ for $M_\chi\gtrsim 0.1$~MeV, with the strongly interacting regime excluded by LHC and current or future direct detection experiments.

\begin{figure*}[!htbp]
\begin{center}
{\includegraphics[width=\columnwidth]{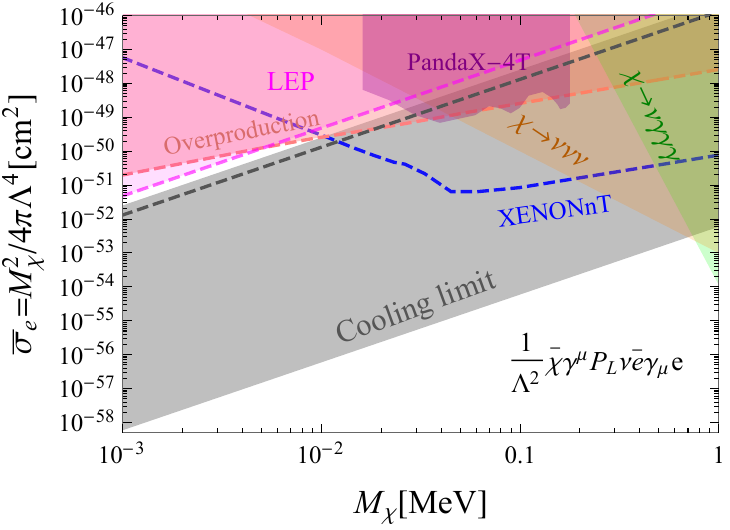}}
{\includegraphics[width=\columnwidth]{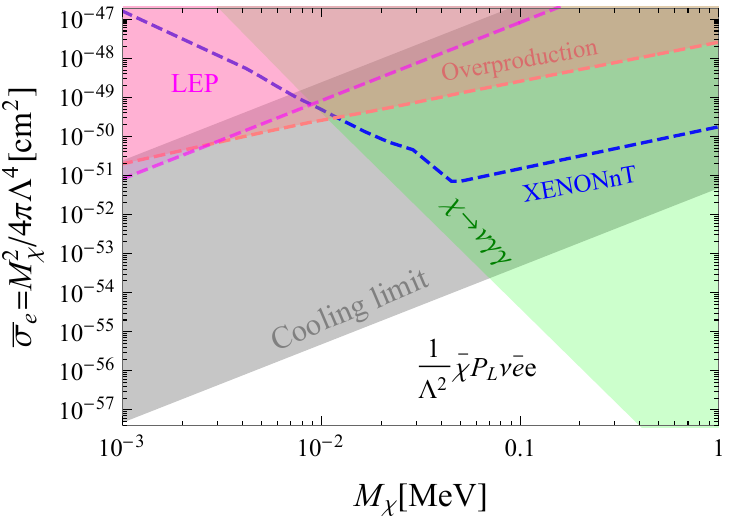}}
\caption{Supernova cooling limits (gray shaded regions) for DM-electron scattering via vector-type (left) and scalar-type (right) interactions. The purple region is excluded by PandaX-4T for fermionic DM absorption on electrons~\cite{PandaX:2022ood,PandaX:2024cic}, and the blue dashed lines are the projected limits of XENONnT experiment extended from~\cite{Dror:2020czw}. Constraints from DM decay are shaded in the upper right~\cite{Dror:2020czw,Ge:2022ius}, and the constraints from DM overproduction via freeze-in are above the pink dashed lines~\cite{Ge:2022ius}. The magenta regions are the constraints from LEP mono-photon searches~\cite{Fox:2011fx}. In the left, we also show the trapping limit when considering the full time information depicted by the dashed gray line.}
\label{fig:coolinglimit}
\end{center}
\end{figure*}

\textbf{\textit{Model setup}}---
FDMCN can be studied both from the UV theory and using effective field theory as a simplified but general framework~\cite{Ge:2022ius}. The relevant degrees of freedom include SM particles, namely, the electron (or nucleon) and the neutrino, along with an additional fermionic DM particle. As no colored degrees of freedom are involved and electroweak symmetry is spontaneously broken, only the electromagnetic $U(1)_{\text{EM}}$ symmetry remains relevant. Consequently, all effective operators involving these particles can be considered, provided they respect $U(1)_{\text{EM}}$ invariance and have appropriate Lorentz structures.

The leading order interactions for FDMCN are
dimension-six operators. We consider two scenarios: DM interacts with electron and DM interacts with nucleon with isospin-independent coupling. For the former, the most commonly studied operators are vector- and scalar-type operators given by
\begin{align}
&O_V =\frac{1}{\Lambda^2}\left(\bar{\chi} \gamma^\mu P_{L} \nu\right)\left(\bar{e} \gamma_\mu e\right)\,,
\label{eq:vectoroperator}\\
&O_S =\frac{1}{\Lambda^2}\left(\bar{\chi} P_{L} \nu\right)(\bar{e} e)\,,
\label{eq:scalaroperator}
\end{align}
where $1/\Lambda^2$ is the Wilson coefficient determined by the heavy mediator mass and its coupling in the UV complete model. The neutrino field is taken to be the SM left-handed component. The latter can be written similarly by replacing the electron with a nucleon in the operators.

Such effective operators are naturally realized in the UV models as discussed in the Supplemental Material~\cite{supp}.  The UV models will not affect the supernova constraints provided that the mediator masses are much larger than the energy of supernova neutrinos; however, they will impact other constraints from DM decay and colliders.

\textbf{\textit{Supernova cooling}}---
Neutrinos are copiously produced in the core of a core-collapse supernova through a variety of processes including neutronization, beta decay and electron-positron annihilation. These neutrinos can scatter with electrons and nucleons on their way of propagation outside the supernova, producing fermionic DM efficiently via the effective interaction in Eq.~\eqref{eq:vectoroperator} and~\eqref{eq:scalaroperator}.
The corresponding total cross sections for neutrino scattering with an electron in the center of mass (COM) frame are given by
\begin{align}
\tilde{\sigma}^V_{\nu e}
&= \frac{x}{12 \pi \sqrt{s}\Lambda^4}
\left[3\tilde{E}_\chi \left(2\tilde{E}_e\tilde{E}_e^{\prime}+\tilde{E}_\nu \tilde{E}_e^{\prime}-m_e^2\right) \right. \notag \\
&\phantom{= \frac{x}{12 \pi \sqrt{s}\Lambda^4} \big[}
\left.+\tilde{p}_f^2\left(3\tilde{E}_e+4\tilde{E}_\nu\right)\right], \label{eq:sigmavector} \\
\tilde{\sigma}^S_{\nu e}
&= \frac{x}{24 \pi \sqrt{s}\Lambda^4}
\left[3\tilde{E}_\chi \left(\tilde{E}_e\tilde{E}_e^{\prime}+m_e^2\right)+ \tilde{p}_f^2\tilde{E}_\nu\right], \label{eq:sigmascalar}
\end{align}
where $x\equiv \tilde{p}_f/\tilde{E}_e$, $\tilde{E}_e$, $\tilde{E}_\nu$, and $\sqrt{s}=\tilde{E}_e+\tilde{E}_\nu$ are the energies of the electron, neutrino and the total energy in the COM, and $\tilde{E}_e^{\prime}=(s-m_\chi^2+m_e^2)/(2\sqrt{s})$ and $\tilde{E}_\chi=(s+m_\chi^2-m_e^2)/(2\sqrt{s})$ are the energies of final state electron and DM in the COM. The final state COM momentum $\tilde{p}_f$ can be inferred from $\tilde{E}_\chi$.

In the supernova, electrons are in thermal and chemical equilibrium with the surrounding matter. We describe the electron energy distribution with the Fermi-Dirac distribution $f_e$ with a space-time dependent temperature $T(t,R)$ and chemical potential $\mu(t,R)$. Since including both the electron and neutrino energy distributions is computationally intensive, we approximate the thermally averaged cross section in Eq.~\eqref{eq:sigmavector} and~\eqref{eq:sigmascalar} by evaluating it at the average electron energy $E_e$ in a head-on collision in the supernova frame.
The DM energy approximately satisfies the relation $E_\nu= (E_\chi+p_\chi)/2$ in this frame. A full kinematic treatment will not affect the results; see Sec.~\ref{sec:average} in the Supplemental Material for details.

The final state electron also experiences a blocking effect as some of the phase space is already occupied by existing electrons with a substantial number density. We take Pauli blocking into consideration by modifying the cross sections with the distribution of electrons, i.e., $\tilde{\sigma}_{\nu e}\rightarrow \sigma_{\nu e}=\tilde{\sigma}_{\nu e}(1-f_e(E'_e,\mu,T))$, where $E'_e$ is the energy of the final state electron in the supernova frame. The dominance of neutrino--electron scattering over electron--positron annihilation is mainly a consequence of the PNS environment rather than the operator structure. For a typical supernova environment~\cite{Manzari:2023gkt}, $n_{\nu_e}/n_{e^+}\simeq2\times10^2$. Because the two channels are crossing-related and have the same Wilson-coefficient dependence, this density hierarchy translates, up to $\mathcal{O}(1)$ kinematic effects, into an approximately two-orders-of-magnitude enhancement of $\nu_e e^-\to\chi e^-$.

The spectrum of fermionic DM produced via the effective DM-neutrino interaction per unit time $t$ at position $r$ is~\cite{Arguelles:2016uwb,Brdar:2023tmi}
\begin{equation}
\frac{1}{4 \pi r^2} \frac{\partial^2}{\partial r \partial t}\left(\frac{d \mathcal{N}_\chi}{d E_\chi}\right)=\sigma_{\nu e} n_e \frac{d n_\nu}{d E_\nu}\,.
\label{eq:differentialeq}
\end{equation}
The neutrino energy spectrum $dn_\nu/dE_\nu=n_\nu f_\nu(E_\nu)/\bar{E}_\nu$ where $\bar{E}_\nu$ is the mean neutrino energy and $f_\nu(E_\nu)$ is the neutrino  distribution function~\cite{Keil:2002in,Brdar:2018zds}
\begin{equation}
f_\nu=\frac{(1+\alpha)^{(1+\alpha)}}{\Gamma(1+\alpha)}\left(\frac{E_\nu}{\bar{E}_\nu}\right)^\alpha \operatorname{Exp}\left[-\left(1+\alpha\right) \frac{E_\nu}{\bar{E}_\nu}\right]\,,
\label{eq:distribution function}
\end{equation}
where $\alpha$ and $\bar{E}_\nu$ are both explicitly functions of time and space. This corresponds to a modified power law distribution with the prefactor being the normalization constant. We work in the regime where the mediator coupling to DM and neutrino is not substantially larger than the mediator-electron coupling, and neglect DM production from neutrino-neutrino scatterings; see the Supplemental Material~\cite{supp} for details.  The opposite regime will also reveal the interaction between DM and neutrinos, which we leave for future work.

We can now substitute the neutrino distribution function in Eq.~\eqref{eq:differentialeq} to obtain the space-time integrated DM flux
\begin{equation}
\begin{aligned}
\frac{d \mathcal{N}_\chi}{d E_\chi} =& \int_0^R 4 \pi R^{\prime 2} d R^{\prime} \int_0^t d t^{\prime} \dfrac{dn_\nu(t',R')}{dE_\nu}\\
& \times\sigma_{\nu e}\left(\Lambda, M_\chi, T\left(t^{\prime}, R^{\prime}\right), E_\nu\right) n_e\left(t^{\prime}, R^{\prime}\right).
\end{aligned}
\label{eq:DMspectrum}
\end{equation}
We assume no flavor dependence in the cross section and sum over all flavor species of neutrinos and antineutrinos.

We extract the space-time dependent variables $\alpha$, $\bar{E}_\nu$, $n_\nu$ and $n_e$, $\mu$, $T$ from an 8.8$M_\odot$ progenitor star simulated by the Garching group~\cite{Hudepohl:2009tyy}, which are illustrated in the Supplemental Material~\cite{supp}. The radial integral extends from the supernova center (close to $R^\prime = 0$) to its outer layers, and we stop at 40 km in our calculation. This cutoff is reasonable since the production of DM predominantly occurs within the neutrino sphere ($\sim$ 30 km), where the number densities of neutrino and electron are orders of magnitude larger than outside. The time integration encompasses the neutronization, accretion, and cooling phases of the supernova explosion, and lasts until 8.85~s after core bounce in the simulation. The neutrino flux is already negligibly small at the end of the cooling phase.

\begin{figure}[!htbp]
\begin{center}
{\includegraphics[width=\columnwidth]{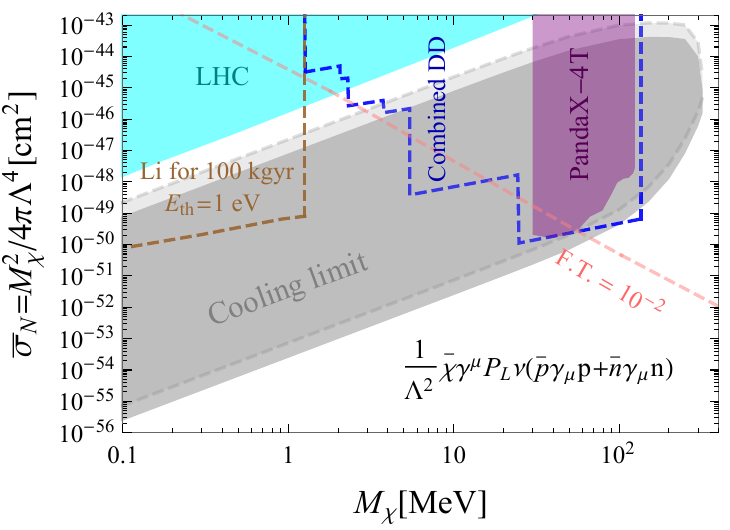}}
\caption{Supernova cooling limits (gray shaded region) for DM-nucleon scattering via vector-type interaction. The purple region is excluded by PandaX-4T for fermionic DM absorption on nucleon~\cite{PandaX:2022osq}, and the blue and brown dashed lines are the combined sensitivities expected from current and future direct detection experiments~\cite{Dror:2019onn,Dror:2019dib}. The cyan region is the constraint from LHC mono-jet searches~\cite{Belyaev:2018pqr,Dror:2019onn}. The level of fine-tuning required to avoid decay constraints is marked by the dashed pink line~\cite{Dror:2019onn}. We have also depicted the cooling limit of scalar-type interaction in the light-gray shaded region enclosed by the dashed line.}
\label{fig:coolinglimitnucleon}
\end{center}
\end{figure}

We then find the total energy loss from DM by integrating out the corresponding spectrum in Eq.~\eqref{eq:DMspectrum}, i.e. $\mathcal{E}_{\chi}=\int E_{\chi}d\mathcal{N}_{\chi}/dE_{\chi}$. The lower boundaries of the constraint regions in Fig.~\ref{fig:coolinglimit} are set using the integrated energy-loss criterion ($\mathcal{E}_\chi<0.1 \mathcal{E}_\nu$), by requiring DM production to carry at most $10\%$ of the energy released in the form of neutrinos computed from the neutrino luminosity extracted from the simulation, which is a common choice in the literature (e.g.~\cite{Dreiner:2013mua,Dreiner:2003wh,Brdar:2023tmi}). Note that we have included the full time evolution information of the supernova rather than the constraint at the luminosity maximum, which yields more robust constraints on DM (See the Supplemental Material~\cite{supp} for comparison with the standard ``Raffelt criterion''~\cite{Raffelt:1990yz}, which yields slightly weaker constraints).
The supernova constraints should therefore be regarded as order-of-magnitude estimates, subject to uncertainties in the progenitor model, nuclear equation of state, neutrino transport, and relevant microphysics.

By applying a similar formalism, the case of DM coupling to nucleons is also considered in Fig.~\ref{fig:coolinglimitnucleon}. We have neglected the Pauli blocking effect of nucleons, the inclusion of which is expected to reduce the production rate by about 20\%; see Sec.~\ref{sec:Pauliblocking} in the Supplemental Material for details.

\textbf{\textit{Dark matter trapping}}---
We now estimate the onset of trapping and identify where free-streaming bounds cease to apply for the cooling constraints. For simplicity, the fully diffusive regime is not modeled. As the $e\nu\to e\chi$ scattering cross section increases, more DM will be produced and leads to larger energy loss. The neutrinos will not be depleted by converting to DM as they are  in thermal and chemical equilibrium with the environmental plasma and will be rapidly replenished. If $1/\Lambda^2$ is too large, the produced DM particle may further scatter with the electron or nucleon to convert back to a neutrino. The back-converted neutrinos may also scatter and produce DM again, contributing to excessive cooling. Conservatively, we neglect this effect and treat DM as lost after scattering. We also neglect the effect of DM-neutrino scattering and DM self-interaction in DM propagation as the number densities of DM and neutrino are much lower than that of electrons~\cite{Fiorillo:2024upk}\footnote{A full transport calculation, including repeated $\chi\leftrightarrow\nu$ conversions and diffusion, would provide a more complete treatment of the trapping regime.}.

The survival probability of DM from back-conversion is
\begin{equation}
P\left(t, r\right)=\mathrm{Exp} \left(-\int_{r}^{\infty} \frac{\mathrm{d} r'}{\lambda\left(t,r'\right)}\right),
\label{eq:escape probability}
\end{equation}
where $\lambda\left(t,r\right)$ is the mean free path of DM. For electron scattering, $\lambda_{\chi e}\left(r\right)=(n_e\sigma_{\chi e \rightarrow \nu e})^{-1}$, and the cross section has been corrected by including the Pauli blocking effect. The mean free path for nucleon scattering can be obtained in a similar way. Here we assume that DM particles produced at radius $r$ travel on radial trajectories out of the supernova. The energy of  DM produced at $(t, R)$ can be estimated as $\bar{E}_\chi=\int\frac{d \mathcal{N}_\chi}{d E_\chi}E_\chi dE_\chi/\int\frac{d \mathcal{N}_\chi}{d E_\chi} dE_\chi$ using Eq.~\eqref{eq:DMspectrum}. We then multiply the survival probability by the integrand of Eq.~\eqref{eq:DMspectrum} and integrate over space and time to compute the trapping-corrected energy loss by DM. The trapping limits can be obtained when the interaction is strong enough so that the integrated energy-loss criterion is again met. We have checked explicitly that assuming random propagation direction instead of radial trajectory will lead to modification in the trapping limits by at most a factor of 3 (see Sec.~\ref{sec:propagationeffects} in the Supplemental Material for detailed discussions).

In principle, the interaction length $\lambda$ is a function of time and radius. However, including both information is time-consuming. We tackle this in two ways:
\begin{itemize}
    \item In the first approach, we fix the time variable and evaluate $\lambda$ at $t=1$~s. This is justified as $\lambda$ is highly correlated with the electron or nucleon number density, which does not vary much at different times in the first 15~km where the number density is highest. The limits are depicted by the gray regions in Fig.~\ref{fig:coolinglimit} and Fig.~\ref{fig:coolinglimitnucleon}.
    \item In the second approach, we fix the radius and evaluate $\lambda$ at $r\simeq 10$~km, where the DM luminosity peaks, as shown in the Supplemental Material~\cite{supp}. By explicitly including the time information in the survival probability along with Eq.~\eqref{eq:DMspectrum}, we obtain the trapping limit as the dashed gray line on the left of Fig.~\ref{fig:coolinglimit} for vector-type interaction (similarly for scalar-type interaction, not shown in Fig.~\ref{fig:coolinglimit}). Despite the different approaches, the trapping limits are in close proximity to each other, demonstrating the robustness of the treatment.
\end{itemize}
Altogether, we find supernova cooling sets stringent constraints on FDMCN, which span by about seven orders of magnitude in the cross section, regardless of the target and the type of interactions. It rules out the cross sections as low as $10^{-51}$–$10^{-58}$ cm$^2$ for DM–electron scattering in the keV–MeV mass range, and $10^{-49}$–$10^{-56}$ cm$^2$ for DM–nucleon scattering in the 0.1–100 MeV mass range. The results are robust against the choice of supernova profiles and conservative compared to models with higher progenitor masses~\cite{supp}.

It should be emphasized that the trapping limits we derive here are conservative, which roughly corresponds to the mean-free paths comparable to the radius of the supernova core, and DM particles are only efficiently trapped at larger cross section.

\textbf{\textit{Other constraints}}---
The special feature of FDMCN also enables DM decay to light SM particles. For DM coupling to electrons, we primarily focus on the sub-MeV mass regime where the decay to electrons is prohibited. The possible decay products are odd numbers of neutrinos with potential photon final states in addition to conserve the angular momentum.

We adopt the estimate of decay rates from different channels in~\cite{Dror:2020czw,Ge:2022ius}, which can be leveraged to recast the corresponding gamma~\cite{Essig:2013goa,Liao:2020hds,Krivonos:2020qvl,Ng:2019gch,Gruber:1999yr,Bouchet:2008rp,Bouchet:2011fn} and neutrino~\cite{Gong:2008gi,DeLopeAmigo:2009dc,Audren:2014bca,Poulin:2016nat,FrancoAbellan:2021sxk} constraints from~\cite{Essig:2013goa,FrancoAbellan:2021sxk}, as detailed in the Supplemental Material~\cite{supp}. We show the DM decay limits for DM-electron interactions in Fig.~\ref{fig:coolinglimit}. DM decay places stringent constraints for heavier DM towards MeV, while lighter DM is less constrained. The DM decay limits are inferior to the supernova cooling limits when $M_\chi\lesssim$~MeV for vector-type coupling and when $M_\chi\lesssim 0.1$~MeV for scalar-type coupling. We also show the constraints recast from LEP mono-photon searches using similar operators~\cite{Fox:2011fx}.

FDMCN could be the cosmological DM.
Sub-MeV fermionic DM is typically produced from the freeze-in process, which is assessed in~\cite{Ge:2022ius} by solving the Boltzmann equation. The constraints of DM overproduction are also shown in Fig.~\ref{fig:coolinglimit}. This corresponds to freeze-in production of DM in the process of $e^+e^-\to\chi\nu$, assuming a maximum temperature of 1~MeV. Higher reheating temperature will further strengthen the overproduction constraints. The supernova constraints we obtain, however, are independent of the cosmological history of DM and the abundance of FDMCN as cosmological DM.

Fermionic DM with nucleon coupling may also be subject to stringent decay constraints. Decay can be suppressed by introducing a UV bare mixing between the SM photon and the vector mediator for vector-type interaction~\cite{Hapitas:2021ilr,Dror:2019onn}, and exotic quarks coupling to the scalar for scalar-type interaction. In both cases, a certain level of fine-tuning is required depending on the DM mass.  The vector mediator is also constrained from monojet searches at the LHC~\cite{Belyaev:2018pqr}, also shown in Fig.~\ref{fig:coolinglimitnucleon}, which is however related to the choice of other model parameters~\cite{Dror:2019onn,Dror:2019dib}. As the thermal history of DM in this scenario varies depending on UV physics and the dynamics in the early Universe~\cite{Dror:2019onn}, we do not place concrete limits on the overproduction of DM.

The constraints from supernova cooling can also be confronted with direct detection experiments.
We show the existing constraints from PandaX-4T for vector-type fermionic DM absorption on electrons~\cite{PandaX:2022ood,PandaX:2024cic} in Fig.~\ref{fig:coolinglimit}. Similar constraints are set by CDEX-10 but in a larger cross section beyond the scope of the figure~\cite{CDEX:2024bum}. We also extend the projected limits of XENONnT~\cite{Dror:2020czw} to larger mass ranges by assuming a constant ratio $\sigma_{\chi e}v_\chi/M_\chi$ using the full essence of the cross section with explicit mass dependence. For DM-nucleon interaction, we show the constraints from PandaX-4T for vector-type fermionic DM absorption on nuclear targets~\cite{PandaX:2022osq}. There are also limits of the Majorana Demonstrator~\cite{Majorana:2022gtu},  CDEX-10~\cite{CDEX:2022rxz} and PICO-60~\cite{PICO:2025rku} extending to slightly different mass ranges. The expected constraints from current experiments are projected in~\cite{Dror:2019onn}, such as CRESST~\cite{CRESST:2015txj,CRESST:2017cdd}, DarkSide-50~\cite{DarkSide:2014llq,DarkSide:2018bpj}, SuperCDMS~\cite{SuperCDMS:2014cds}, PICO~\cite{PICO:2015pux,PICO:2017tgi} and xenon experiments~\cite{LUX:2016ggv,PandaX-II:2017hlx,XENON:2018voc}, along with the sensitivity of a future Lithium-target experiment with 100~kg$\cdot$yr exposure for the same type of interaction. Although some of the limits in the literature were obtained assuming coupling to a massless right-handed neutrino, the scattering cross section remains unchanged in comparison with the left-handed ones. The direct detection constraints on scalar-type interaction are expected to be similar, with slight modifications from the different differential cross sections.

\textbf{\textit{Complementarity}}---
The cooling limits have strong complementarity with constraints from DM decay, cosmology, collider and direct detection experiments. For DM coupling to electrons, supernova cooling sets more stringent constraints than PandaX and even the projection from XENONnT. It rules out almost the whole parameter space for fermionic DM to constitute $\mathcal{O}(1)$ fraction of cosmological DM in the keV to MeV mass window, in particular for scalar-type interaction. In combination with the overproduction and collider limits, supernova cooling excludes the DM-electron scattering cross section down to $10^{-51}-10^{-58}$~cm$^2$.
For DM coupling to nucleons, supernova cooling rules out large parameter space below the LHC monojet constraint. The small parameter space in between can be covered by existing or future DM direct detection experiments.

We note that the supernova limits do not require the new particles to constitute cosmological DM, which is a key distinction from direct and indirect detection experiments. The limits can also be scaled and extended to arbitrarily small masses of new particles.

\textbf{\textit{Summary and outlook}}---
The extreme temperature and density environment in the supernova provides a powerful laboratory to create sub-GeV particles in the dark sector, where the neutrino burst in the supernova explosion can be leveraged to study the interactions between neutrinos and DM.

In this \textit{Letter}, we investigate for the first time neutrino-DM conversion, and identify a previously unexplored production channel for fermionic dark matter via the neutrino-devouring process inside a core-collapse supernova. We set stringent and robust constraints on the interaction strength in the sub-MeV (DM-electron coupling) and sub-GeV (DM-nucleon coupling) mass ranges, which can improve upon direct detection limits by several orders of magnitude, and provide strong complementarity with other constraints. These results significantly advance our understanding of DM and DM-neutrino interactions, which are of broad interest to a wide audience in particle physics, astrophysics and cosmology. Our
work can also be generalized to other types of interactions, or UV models with heavy or light mediators.

%%%%%%%%%%%%%%%%%%%%%%%%%%%%%%%%%%%%%%%%%%%%%
\section*{Acknowledgments}
%%%%%%%%%%%%%%%%%%%%%%%%%%%%%%%%%%%%%%%%%%%%%
We thank Ying-Ying Li for assistance with the calculations in this work. We also thank Damiano F.G. Fiorillo, Gang Li, Robert McGehee, and  Edoardo Vitagliano for useful correspondence. NS is supported by the National Natural Science Foundation of China (NNSFC) Project Nos. 12347105, 12475110, 12441504 and 12447101. C.-T. Lu is supported by the NNSFC under grant Nos. 12335005, 12575118 and the Special funds for postdoctoral overseas recruitment, Ministry of Education of China. Y.L. is supported by the NNSFC under grant No. 12441504.

\bibliography{main}

%%%%%%%%%%%%%%%%%%%%%%%%%%%%%%%%%%%%%%%%%%%%%%%%%%%%%%%%%%%%%%%%%%%%%%%%%%%%%%%%%%%%%%%%%%%%
%%%%%%%%%%%%%%%%%%%%%%%%%%%%%%%%%%%%%%%%%%%%%%%%%%%%%%%%%%%%%%%%%%%%%%%%%%%%%%%%%%%%%%%%%%%%

\onecolumngrid
\clearpage

\begin{center}
\textbf{\large Supernova cooling from neutrino-devouring dark matter}

\vspace{0.05in}
{ \it \large Supplemental Material}\\
\vspace{0.05in}
{Yugen Lin, Chih-Ting Lu, and Ningqiang Song}
\end{center}
\onecolumngrid
\setcounter{equation}{0}
\setcounter{figure}{0}
\setcounter{section}{0}
\setcounter{table}{0}
\setcounter{page}{1}
\makeatletter
\vspace{-3mm}

\section{UV complete models}
\label{sec:UVmodels}

The interactions considered in the main text are effective operators below the electroweak scale.
 Here we present explicit gauge-invariant realizations of the scalar and vector interactions. We emphasize that the UV constructions presented here are intended only
as illustrative examples showing that the effective interactions used in
the main text can arise from renormalizable electroweak-gauge-invariant
theories. They are neither unique nor claimed to be minimal.  Other UV
realizations, with different mediator sectors, symmetry assignments, or
heavy-field content, may reproduce the same low-energy operators. Our
purpose is therefore not to provide an exhaustive classification of UV
completions, but to demonstrate explicitly how the interactions involving
left-handed active neutrinos can be embedded consistently above the
electroweak scale and to identify the additional interactions that such
embeddings may entail.

\subsection{Scalar interaction}
\label{sec:scalar-UV}

We now illustrate that the scalar interaction used in the supernova
calculation,
\begin{equation}
    \mathcal{L}_{\rm EFT}^{S}
    =
    \frac{1}{\Lambda^2}
    \left(\bar\chi P_L\nu_e\right)
    \left(\bar e e\right)
    +{\rm h.c.},
\label{eq:scalar_EFT}
\end{equation}
can arise from a renormalizable electroweak-gauge-invariant theory.
Besides the Majorana fermion $\chi$, we introduce a real scalar singlet
$S$, a neutral Dirac fermion $N_{L,R}$, and a charged vector-like lepton
$E_{L,R}\sim({\bf 1},{\bf 1},-1)$. The relevant interactions are
organized by two approximate parities,
\begin{center}
\begin{tabular}{c|ccccccc}
 & $L_e$ & $H$ & $e_R$ & $N_{L,R}$ & $E_{L,R}$ & $\chi_R$ & $S$ \\
\hline
$Z_2^X$ & $+$ & $+$ & $-$ & $+$ & $+$ & $-$ & $-$ \\
$Z_2^D$ & $+$ & $+$ & $+$ & $-$ & $+$ & $-$ & $+$
\end{tabular}
\end{center}
The SM electron Yukawa $y_e\bar L_eHe_R$ and the active--heavy
Yukawa $y_N\bar L_e\widetilde HN_R$ are treated as the small
$Z_2^X$- and $Z_2^D$-breaking spurions, respectively. Here
$\widetilde H=i\sigma_2H^\ast$. We also assume approximate lepton
number, with $N$, $E$, and $\chi$ carrying lepton number one and the
Majorana mass $m_\chi$ providing the displayed $\Delta L=2$ breaking.

The renormalizable interactions relevant for matching are
\begin{align}
-\mathcal{L}_{\rm UV}\supset {}&
 M_N\bar N_LN_R
 +M_E\bar E_LE_R
 +\frac{1}{2}m_\chi\overline{\chi^c}\chi
 +y_e\bar L_eHe_R
 +y_N\bar L_e\widetilde HN_R
 \nonumber\\
&+\lambda_E\bar L_eHE_R
 +y_D S\bar N_L\chi_R
 +y_E S\bar E_Le_R
 +{\rm h.c.}.
\label{eq:single_scalar_UV}
\end{align}
After electroweak symmetry breaking, we define
\begin{equation}
    \theta_N
    \equiv
    \frac{y_Nv}{\sqrt{2}M_N},
    \qquad
    \theta_E
    \equiv
    \frac{\lambda_Ev}{\sqrt{2}M_E},
\label{eq:theta_NE}
\end{equation}
where $\theta_N$ is the $\nu_e$--$N$ mixing angle and $\theta_E$
is the $e$--$E$ mixing angle. To leading order in these small
mixings,
\begin{equation}
-\mathcal{L}
\supset
g_\chi S\,\bar\chi P_L\nu_e
+
g_e S\,\bar e e
+{\rm h.c.},
\qquad
g_\chi=y_D\theta_N,
\qquad
g_e=y_E\theta_E .
\label{eq:low_scalar_couplings}
\end{equation}
Integrating out $S$ therefore gives
\begin{equation}
    \frac{1}{\Lambda^2}
    =
    \frac{g_\chi g_e}{m_S^2}
    =
    \frac{y_Dy_E\theta_N\theta_E}{m_S^2}.
\label{eq:single_scalar_matching}
\end{equation}

As an explicit perturbative benchmark, we take
\begin{align}
&m_\chi=5~{\rm keV},
\qquad
m_S=2~{\rm GeV},
\qquad
M_N=100~{\rm GeV},
\qquad
M_E=1.5~{\rm TeV},
\nonumber\\
&y_D=0.2,
\qquad
y_E=1,
\qquad
\theta_N=5.0\times10^{-5},
\qquad
\theta_E=4.0\times10^{-5}.
\label{eq:single_scalar_benchmark}
\end{align}
This corresponds to
\begin{equation}
    y_N\simeq2.9\times10^{-5},
    \qquad
    \lambda_E\simeq3.4\times10^{-4},
\end{equation}
and gives
\begin{equation}
    g_\chi\simeq1.0\times10^{-5},
    \qquad
    g_e\simeq4.0\times10^{-5},
    \qquad
    \Lambda\simeq100~{\rm TeV}\,,
\end{equation}
roughly within the parameter range considered in this work.

We now consider other ingredients and constraints associated with the model.
\paragraph{Scalar and Higgs sector.}
In the exact $Z_2^X$ limit the scalar potential can be written as
\begin{equation}
    V_0
    =
    V_H
    +\frac{1}{2}m_{S0}^2S^2
    +\frac{\lambda_S}{4!}S^4
    +\frac{\lambda_{HS}}{2}S^2H^\dagger H ,
\label{eq:single_scalar_potential}
\end{equation}
where $V_H=-\mu_H^2H^\dagger H+\lambda_H(H^\dagger H)^2$.
Terms containing an odd number of $S$ fields are $Z_2^X$ odd.
Once the electron-Yukawa spurion is included, the renormalized theory
therefore also contains
\begin{equation}
    V_{\rm br}
    =
    t_S S
    +a_S S H^\dagger H
    +\frac{\kappa_S}{3!}S^3 ,
\label{eq:single_scalar_breaking}
\end{equation}
with coefficients proportional to the $Z_2^X$-breaking spurion.
Loop-induced scalar tadpoles are included in the usual renormalized
vacuum condition.

The linear portal $a_S S H^\dagger H$ generates Higgs--singlet mass
mixing. Since it is $Z_2^X$ odd, it is absent in the symmetry limit
and is spurion suppressed. Parametrically, the charged-lepton sector
generates
\begin{equation}
    a_S^{\rm loop}
    \sim
    \frac{y_e\,y_E\,\lambda_E}{16\pi^2}M_E ,
\end{equation}
up to an order-one loop function. For the benchmark above this gives
a very small Higgs--singlet mixing angle,
\begin{equation}
    \theta_{hS}
    \simeq
    \frac{a_Sv}{m_h^2-m_S^2}
    \lesssim {\cal O}(10^{-10}),
\end{equation}
so Higgs-induced couplings of $S$ to quarks and nucleons are negligible.

The quadratic Higgs portal $\lambda_{HS}$ is allowed by the discrete
symmetries and is therefore not symmetry protected. Its radiative
size is nevertheless small for the displayed benchmark,
\begin{equation}
    \delta\lambda_{HS}
    \sim
    \frac{y_E^2\lambda_E^2}{16\pi^2}
    +
    \frac{y_D^2y_N^2}{16\pi^2}
    \sim {\cal O}(10^{-9}).
\end{equation}
Thus an illustrative value such as $\lambda_{HS}=10^{-5}$ is
radiatively stable against the displayed portal interactions. It
gives
\begin{equation}
    \Gamma(h\to SS)
    =
    \frac{\lambda_{HS}^2v^2}{32\pi m_h}
    \sqrt{1-\frac{4m_S^2}{m_h^2}}
    \simeq0.48~{\rm eV},
\end{equation}
corresponding to
\begin{equation}
    {\rm BR}(h\to SS)\simeq1.2\times10^{-7}.
\end{equation}
Since the mediator is an electroweak singlet, the completion also
does not contain the light charged-scalar partner characteristic of
an inert-doublet realization, and hence avoids the associated
nondecoupling charged-scalar contribution to $h\to\gamma\gamma$.

\paragraph{Active--DM mixing and radiative operators.}
Gauge invariance permits the direct operator $y_{\chi H}\bar L_e\widetilde H\chi_R$. In the present construction it is odd under both $Z_2^X$ and $Z_2^D$.
Consequently, if $y_e$ and $y_N$ are the only breaking spurions,
\begin{equation}
    y_{\chi H}
    \propto
    y_e y_N
    \times
    \left(\text{portal and loop factors}\right).
\label{eq:LHchi_spurion}
\end{equation}
The smallness of the active--DM mixing is therefore technically
organized by the two approximate symmetries. Scalar tadpoles generated
after $Z_2^X$ breaking are included and renormalized together with the
neutral-fermion mixing parameters.

The same two-spurion selection rule applies to the gauge-invariant
electroweak dipole operators
\begin{align}
    {\cal O}_B
    &=
    \left(\bar L_e\sigma^{\mu\nu}\chi_R\right)
    \widetilde H\,B_{\mu\nu},
\label{eq:dipoleB}\\
    {\cal O}_W
    &=
    \left(\bar L_e\sigma^{\mu\nu}\tau^I\chi_R\right)
    \widetilde H\,W^I_{\mu\nu}.
\label{eq:dipoleW}
\end{align}
Both operators are odd under $Z_2^X$ and $Z_2^D$ and therefore require
both symmetry-breaking spurions,
\begin{equation}
    C_B,\;C_W
    \propto
    y_e y_N
    \times
    \left(\text{portal and loop factors}\right).
\end{equation}
A complete
on-shell matching calculation determines the finite coefficients, but
the symmetry makes the dipole
operators are parametrically suppressed.

\paragraph{Relation to sterile-neutrino and inert-doublet models.}
The neutral fermion $N$ plays the role of a sterile-neutrino-like
transition state, while approximate lepton number keeps it Dirac (or
pseudo-Dirac) and suppresses radiative Majorana neutrino masses.  The
construction is therefore related to sterile-neutrino portal models,
but differs from the minimal inert-doublet or scotogenic realization.
In particular, the light scalar mediator is an electroweak singlet,
and its electron coupling is generated through the heavy charged
vector-like lepton $E$. The model consequently does not require a
light electroweak scalar doublet or a light charged scalar.

For the benchmark above,
\begin{equation}
    |U_{eN}|^2\simeq\theta_N^2=2.5\times10^{-9},
    \qquad
    \theta_E=4.0\times10^{-5},
\end{equation}
so the mixings of the additional fermions with the SM leptons are
small. The heavy states are not thermally populated in the
proto-neutron-star core, while their virtual effects reproduce the
low-energy interaction in Eq.~\eqref{eq:scalar_EFT}.

\subsection{Vector interaction}
\label{sec:vector-UV}

We also give an explicit renormalizable vector completion of the interaction used in the supernova analysis. The purpose is to make explicit how a current containing the active left-handed neutrino can arise from a gauge theory, and which additional interactions necessarily accompany it.

\paragraph{Gauge structure and renormalizable interactions.}
We extend the gauge group to $G_{\rm SM}\times SU(2)_D\times U(1)_X$. The SM fields are neutral under the dark gauge group $SU(2)_D\times U(1)_X$.  The complete new field content consists of two
left-handed dark fermion doublets $\Psi_L,\ \Psi'_L\sim({\bf 2},0)$,
four right-handed dark fermion singlets
    $N_R,\ \chi_R,\ \xi_R,\ \eta_R
    \sim({\bf 1},0)$,
and three dark scalars,
    $\Phi\sim({\bf 2},0)$,
    $\Omega\sim({\bf 2},1)$,
    $S\sim({\bf 1},1)$,
under $SU(2)_D\times U(1)_X$;
all of these fields are singlets under the SM gauge group. The electron
portal is mediated by two pairs of vector-like leptons,
    $E_{\pm,L/R}\sim({\bf 1},{\bf 1},-1)$,
    $D_{\pm,L/R}\sim({\bf 1},{\bf 2},-1/2)$,
under $SU(3)_c\times SU(2)_L\times U(1)_Y$.  They are singlets under
$SU(2)_D$ and carry $U(1)_X$ charges $X=\pm1$, with the subscript
$\pm$ denoting the dark $U(1)_X$ charge rather than the electric charge. The vacuum expectation values
\begin{equation}
\langle\Phi\rangle=\frac{v_D}{2}(1,1)^T,\qquad
\langle\Omega\rangle=\frac{v_\Omega}{\sqrt2}(0,1)^T,\qquad
\langle S\rangle=\frac{v_X}{\sqrt2}
\end{equation}
break $SU(2)_D\times U(1)_X$ completely. The VEV of $\Phi$ generates
the Dirac masses $m_N=y_N^D v_D/\sqrt2$ and
$m_\chi=y_\chi^D v_D/\sqrt2$, $v_\Omega$ induces mixing between the
neutral $SU(2)_D$ and $U(1)_X$ gauge bosons, and $v_X$ generates the
electron--vector-like-lepton portal mixings. The dark gauge couplings are denoted by $g_D$ and $g_X$, corresponding
to $SU(2)_D$ and $U(1)_X$, respectively.
The VEVs $v_D$, $v_\Omega$, and $v_X$ generate the masses of the dark
gauge bosons. In particular, in the neutral $(W_D^3,X)$ basis,
\begin{equation}
    \mathcal M_V^2
    =
    \begin{pmatrix}
    \dfrac{g_D^2(v_D^2+v_\Omega^2)}{4}
    &
    -\dfrac{g_Dg_Xv_\Omega^2}{2}
    \\[2mm]
    -\dfrac{g_Dg_Xv_\Omega^2}{2}
    &
    g_X^2(v_X^2+v_\Omega^2)
    \end{pmatrix}.
\label{eq:dark_vector_mass}
\end{equation}
The off-diagonal entry generated by $v_\Omega$ mixes $W_D^3$ and
$X_\mu$, so the neutral vector mass eigenstates couple simultaneously
to the dark transition current and to the electron current.

The fields $\Psi_L$, $N_R$, and $\chi_R$ generate the neutral
$\nu_e$--$\chi$ transition current, while $\Psi'_L$, $\xi_R$, and
$\eta_R$ complete the dark fermion sector and remove otherwise isolated
spectator numbers. The two chiral $SU(2)_D$ doublets $\Psi_L$ and
$\Psi'_L$ cancel the $SU(2)_D$ Witten anomaly, while the SM-charged
fermions $E_\pm$ and $D_\pm$ are vector-like under all gauge factors, so
the perturbative gauge anomalies cancel.
An approximate global $Z_5$ symmetry is used as an organizing symmetry for the dark Yukawa texture. A convenient charge assignment is
 $q_5(\Psi_L,\Psi'_L,N_R,\chi_R,\xi_R,\eta_R,\Phi,\Omega,S)
 =(1,4,0,2,3,0,1,0,0)$,
while the vector-like electron-portal fermions are neutral under $Z_5$.

The interactions responsible for the neutral transition can be written as
\begin{align}
-\mathcal L_{\rm n}\supset{}&
 y_N^D\,\bar\Psi_L\Phi N_R
 +y_\chi^D\,\bar\Psi_L\widetilde\Phi\chi_R
 +y_\nu\,\bar L_e\widetilde H N_R
 +{\rm h.c.},
\label{eq:vector_neutral_yukawa}
\end{align}
where $\widetilde\Phi=i\sigma_2\Phi^*$ and $\widetilde H=i\sigma_2H^*$. The $Z_5$ charge assignment forbids the direct operator
    $\bar L_e \widetilde H \chi_R$
at the tree-level boundary, while allowing
    $y_\nu \bar L_e \widetilde H N_R$.
Consequently, the active--heavy mixing required to generate the
$\nu_e$--$\chi$ transition current does not induce a tree-level
active--DM mass mixing. The electron portal is
\begin{align}
-\mathcal L_{\rm VL}\supset{}&
\sum_{s=\pm}
\left(
 M_{E_s}\bar E_{sL}E_{sR}
 +M_{D_s}\bar D_{sL}D_{sR}
\right)
\nonumber\\
&+
\lambda_+ S\left(\bar E_{+L}e_R+\bar D_{+R}L_e\right)
+
\lambda_- S^*\left(\bar E_{-L}e_R+\bar D_{-R}L_e\right)
+{\rm h.c.}
\label{eq:vector_electron_portal}
\end{align}
Together with the dark gauge interactions, Eqs.~\eqref{eq:vector_neutral_yukawa} and \eqref{eq:vector_electron_portal} are manifestly invariant under the SM gauge group and $SU(2)_D\times U(1)_X$.

After symmetry breaking, $y_\nu$ generates an active--heavy neutral mixing
\begin{equation}
    \theta_N\simeq
    \frac{y_\nu v}{\sqrt2\,M_N} \, ,
\label{eq:vector_thetaN}
\end{equation}
where $M_N$ denotes the predominantly heavy neutral-fermion mass. Importantly, $\theta_N$ is the $\nu_e$--$N$ mixing and is not a tree-level $\nu_e$--$\chi$ mass mixing.  The $S$ expectation value similarly mixes the electron with the $X$-charged vector-like leptons, so that the dark vectors acquire an electron current after diagonalizing the charged-fermion mass matrix.

The dark scalar expectation values generate a transition gauge current between the predominantly $N$ and $\chi$ states. The VEV $v_\Omega$ mixes the
neutral gauge bosons $W_D^3$ and $X_\mu$ into two mass eigenstates
$V_{1,2}$ with mixing angle $\alpha_D$. Denoting the four relevant dark-vector mass eigenstates by $V_a$, their mass-basis currents contain
\begin{equation}
    \mathcal L\supset
    V_{a\mu}
    \left[
       g^a_{\chi\nu}\,\bar\chi\gamma^\mu P_L\nu_e
       +\bar e\gamma^\mu
       \left(g^a_{eL}P_L+g^a_{eR}P_R\right)e
       +\cdots
    \right] .
\end{equation}
The $\chi$--$\nu_e$
transition current originates from the $SU(2)_D$ interaction and is,
to leading order, $g_{\chi\nu}\simeq \frac{g_D}{2}\theta_N$. The electron
inherits a $U(1)_X$ coupling through mixing with the vector-like leptons, $g_{eV}=(g_{eL}+g_{eR})/2\simeq
    g_X\left(\epsilon_+^2-\epsilon_-^2\right)$,
    $\epsilon_\pm\equiv
    \frac{\lambda_\pm v_X}{\sqrt{2}M}$.
Consequently, integrating out $V_{1,2}$ gives approximately
\begin{equation}
    \frac{1}{\Lambda^2}
    \simeq
    \frac{g_Dg_X}{4}\,
    \theta_N
    \left(\epsilon_+^2-\epsilon_-^2\right)
    \sin 2\alpha_D
    \left(
        \frac{1}{m_{V_1}^2}
        -
        \frac{1}{m_{V_2}^2}
    \right),
\label{eq:vector_matching}
\end{equation}
up to an overall sign convention.

As an explicit tree-level benchmark, we take
\begin{equation}
 (v_D,v_\Omega,v_X)=(60,63.2,48)~{\rm GeV},
 \qquad
 (g_D,g_X)=(0.1,0.25),
\end{equation}
with a predominantly heavy neutral state of mass $M_N\simeq50$ GeV, $m_\chi=5$ keV, and vector-like leptons near $1$ TeV. The active--heavy entry is $y_\nu v/\sqrt2=0.05$ GeV, giving $\theta_N\simeq10^{-3}$. The light dark-fermion mass is fixed by
\begin{equation}
    m_\chi=\frac{y_\chi^D v_D}{\sqrt2}=5~{\rm keV},
    \qquad
    y_\chi^D=1.18\times10^{-7}.
\end{equation}
Exact tree-level diagonalization gives dark-vector masses $m_{V_a}\simeq(4.36,\,4.36,\,3.53,\,20.0)~{\rm GeV}$,
and $\Lambda\simeq100$ TeV. Thus the required left-handed-neutrino interaction can be obtained without identifying the active--heavy angle $\theta_N$ with a physical active--DM mixing angle.

\paragraph{Dark-scalar and Higgs sector.}
A compact scalar potential that realizes the required dark vacuum is obtained by defining
 $p=\Phi^\dagger\Phi$,
 $o=\Omega^\dagger\Omega$,
 $z=|S|^2$,
 $t=H^\dagger H-\frac{v^2}{2}$,
 $I=\Phi^\dagger\Omega$,
 $J=\Phi^T i\sigma_2\Omega$,
and taking
\begin{align}
V={}&
 \lambda_H t^2
 +\lambda_p(p-p_0)^2
 +\lambda_o(o-o_0)^2
 +\lambda_z(z-z_0)^2
\nonumber\\
&+\kappa\left(|I-kS|^2+|J-kS|^2\right)
 +\epsilon\,t\left[(p-p_0)+(o-o_0)+(z-z_0)\right].
\label{eq:vector_scalar_potential}
\end{align}
For
 $p_0=\frac{v_D^2}{2}$,
 $o_0=\frac{v_\Omega^2}{2}$,
 $z_0=\frac{v_X^2}{2}$,
this potential has the desired classical vacuum and a positive physical scalar Hessian after the four dark Goldstone modes are removed. The Higgs--dark-sector portal is controlled by $\epsilon$ and is not required to obtain the matching in Eq.~\eqref{eq:vector_matching}. Taking the illustrative boundary value $\epsilon=10^{-6}$ gives only small Higgs--dark-scalar mixing. At the displayed benchmark, the calculated new Higgs two-body width corresponds to ${\rm BR}_{\rm new}(h)\simeq3.5\times10^{-5}$,
so Higgs exotic decays do not appreciably modify the parameter point.

\paragraph{Active--DM mixing and radiative operators.}
The neutral mass matrix at the benchmark contains no tree-level $\nu_e$--$\chi$ mass entry.
At the electroweak-symmetric level, the transition to a photon is described by the gauge-invariant dipole operators similar to Eqs.~\eqref{eq:dipoleB} and~\eqref{eq:dipoleW}.
They are not present as renormalizable tree-level interactions in the UV Lagrangian. Their coefficients arise only after loop matching of the complete theory.

\section{Supernova energy loss and simulation Data}
\label{sec:simulationdata}

\begin{figure}[!htb]
\includegraphics[width=0.49\columnwidth]{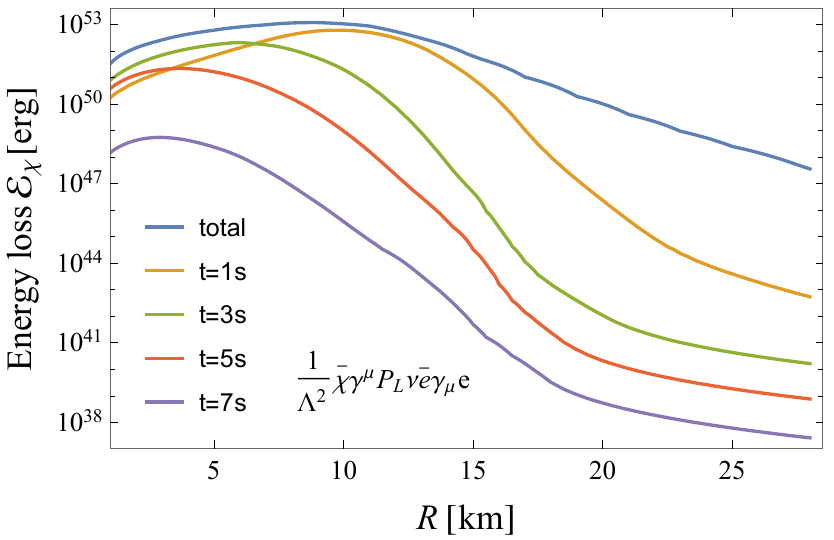}
\includegraphics[width=0.49\columnwidth]{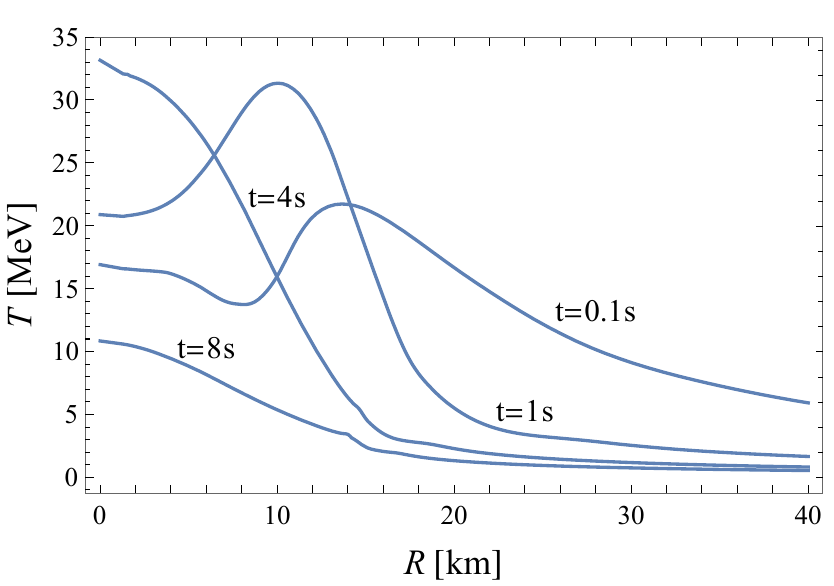}
\caption{{\it Left:} Supernova energy loss from fermionic DM production in  1-km-wide spherical shells and 1 s time slides as a function of radius, i.e. $4\pi R^2 \frac{d^3\mathcal{N}_\chi}{dE_\chi dRdt}\Delta R\Delta t$ with $\Delta R=1$~km and $\Delta t=1$~s. The total energy loss is the time-integrated energy loss in the shells. Here we assume a vector-type coupling to electrons with $M_\chi=1$ MeV and $\Lambda^{-1}=10^{-7}~\rm{MeV}^{-1}$ for illustration, while other interactions have similar behavior. {\it Right:} The profile of supernova temperature.}
\label{fig:Eloss_chi}
\end{figure}

\begin{figure*}[!htb]
\centering
\includegraphics[width=0.49\columnwidth]{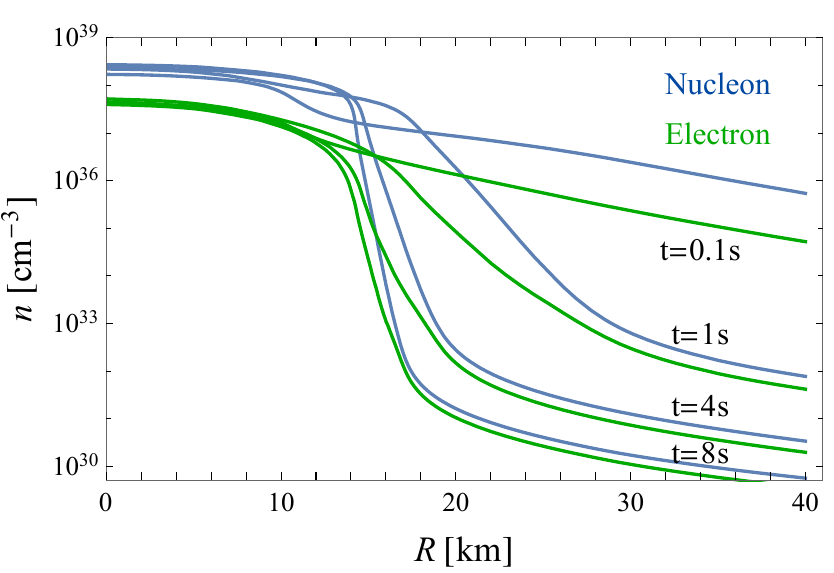}
\includegraphics[width=0.49\columnwidth]{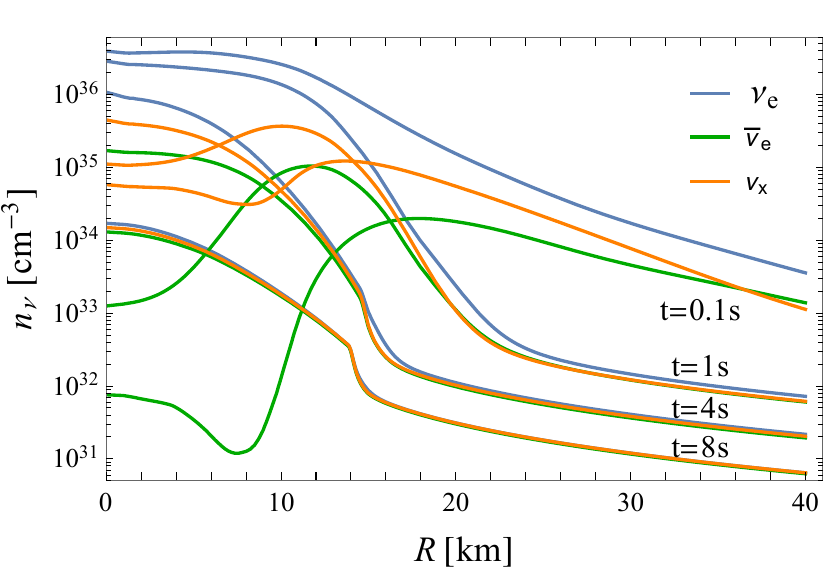}
\includegraphics[width=0.49\columnwidth]{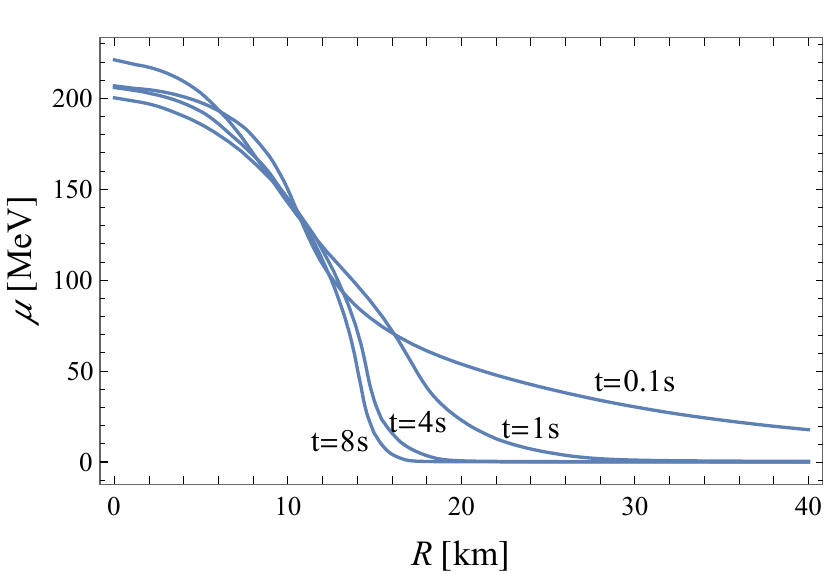}
\includegraphics[width=0.49\columnwidth]{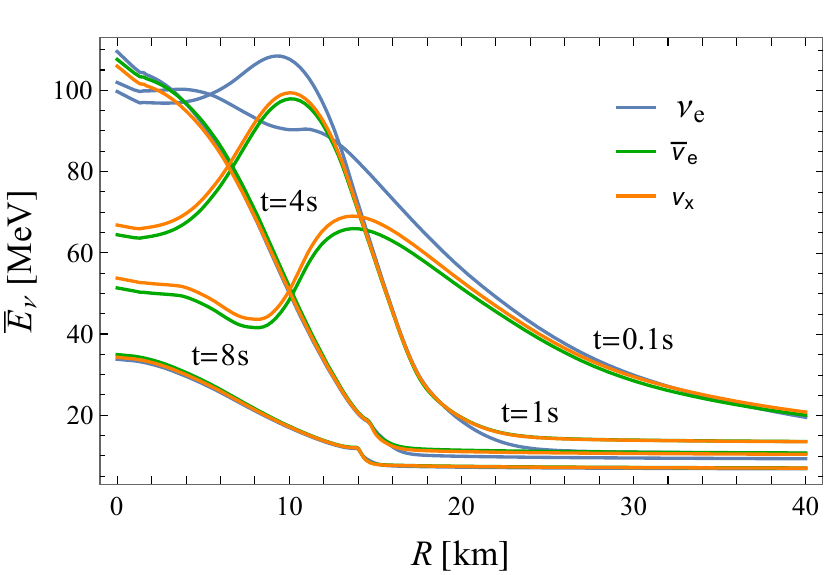}

\caption{ The profile of baryons $n_N$, electrons $n_e$, neutrinos $n_\nu$ number densities and electron chemical potential $\mu$, mean energy of neutrino $\bar{E}_\nu$, and temperature $T$ from supernova simulaitons at different radii for three typical timescales $t=$ 0.1~s, 1~s, 4~s, and 8~s.
{\it Top left:} The profile of the baryon and electron number density. {\it Top right:} The profile of the neutrino number density, including $\nu_e$, $\bar{\nu}_e$ and the sum of other neutrino flavors $\nu_x$. {\it Bottom left:} The profile of the electron chemical potential. {\it Bottom right:} The profile of mean neutrino energy. }
\label{fig:simulationdata}
\end{figure*}

In Fig.~\ref{fig:Eloss_chi}
we show the energy loss from fermionic DM. The energy loss roughly peaks at 10~km.  We also show the temperature profile of the supernova for three typical timescales $t=$ 1~s, 4~s, and 8~s after core bounce from the simulation data of an 8.8$M_\odot$ progenitor star~\cite{Hudepohl:2009tyy}.

In Fig.~\ref{fig:simulationdata}, we show the number densities of baryons $n_N$, electrons $n_e$, neutrinos $n_\nu$ , and electron chemical potential $\mu$, mean energy of neutrino $\bar{E}_\nu$ at different radii from supernova simulations.

\section{Constraints from dark matter decay}

As motivated by the UV models in Sec.~\ref{sec:UVmodels}, the DM decay channels can be considered from the symmetry point of view. We first consider DM coupling to electrons. The decay generally happens at the loop level. For vector-type interaction, the decay channel $\chi\rightarrow \nu+\gamma$ vanishes due to gauge symmetry and $\chi\rightarrow \nu+\gamma\gamma$ vanishes due to the charge conjugation symmetry. The channels $\chi\rightarrow \nu\nu\nu$ and $\chi\rightarrow \nu+\gamma\gamma\gamma$ are allowed through an electron loop.

For scalar-type interaction, the decay modes involving odd numbers of photons $\chi\rightarrow \nu+\gamma$ and $\chi\rightarrow \nu+\gamma\gamma\gamma$ vanish due to the charge conjugation symmetry. The decay $\chi\rightarrow \nu\nu\nu$ also vanishes considering the structure of operator~\cite{Ge:2022ius}. Therefore, the leading decay channel is $\chi\rightarrow \nu+\gamma\gamma$.

DM decay involving photons will result in X-ray or gamma ray excess observed by telescopes such as NuSTAR and INTEGRAL~\cite{Essig:2013goa,Liao:2020hds,Krivonos:2020qvl,Ng:2019gch,Gruber:1999yr,Bouchet:2008rp,Bouchet:2011fn}, and DM decay to neutrinos will increase the effective degree of freedom of the Universe which is constrained by the Cosmic Microwave Background~\cite{Gong:2008gi,DeLopeAmigo:2009dc,Audren:2014bca,Poulin:2016nat,FrancoAbellan:2021sxk}. We adopt the estimate of decay rates from different channels in~\cite{Dror:2020czw,Ge:2022ius}, which can be leveraged to recast the corresponding gamma and neutrino constraints from~\cite{Essig:2013goa,FrancoAbellan:2021sxk}.

Fermionic DM with nucleon coupling may also be subject to stringent decay constraints. For vector-type coupling, quark loops introduce an effective kinetic mixing $\epsilon$ between $A'$ and SM photon, which enables the decay mode $\chi\rightarrow \nu+e^+e^-$ ($\mu^+\mu^-$) in the sub-GeV mass range, and $\chi\rightarrow \nu+\gamma\gamma\gamma$ in the sub-MeV mass range. As with electrons, they are expected to exclude the cross section for heavier mass but less so for lighter DM. The kinetic mixing could be suppressed by introducing a UV bare mixing $\epsilon_{\rm UV}$~\cite{Hapitas:2021ilr}. The decay rate is then determined by the net kinetic mixing through the fine-tuning parameter F.T.$=|\epsilon_{\rm UV}-\epsilon|/|\epsilon|$, and the decay is largely avoided when F.T. is small enough~\cite{Dror:2019onn}.

For DM coupling to nucleons via a scalar-type interaction, DM decay may occur through $\chi\rightarrow \nu+\gamma\gamma$ with a quark loop. The decay could be suppressed by introducing exotic quarks coupling to the scalar with an opposite sign; however, a similar level of fine-tuning is required.

\section{Comparison between energy-loss criterion and Raffelt criterion}

\begin{figure}[!htb]
\includegraphics[width=0.49\columnwidth]{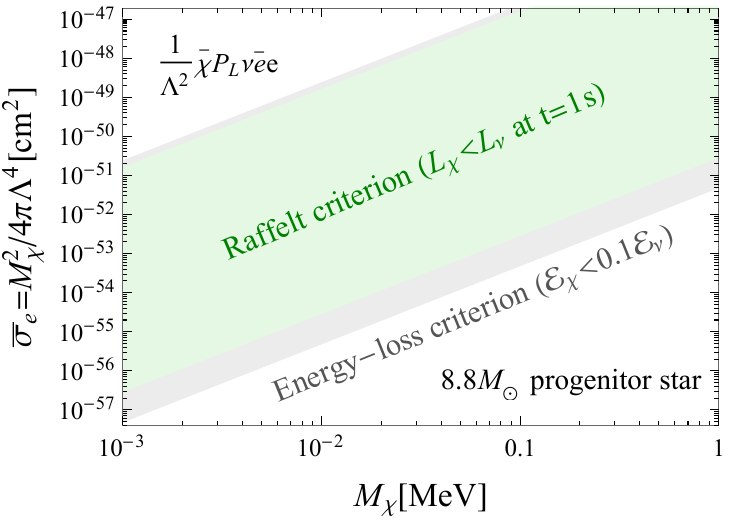}
\includegraphics[width=0.49\columnwidth]{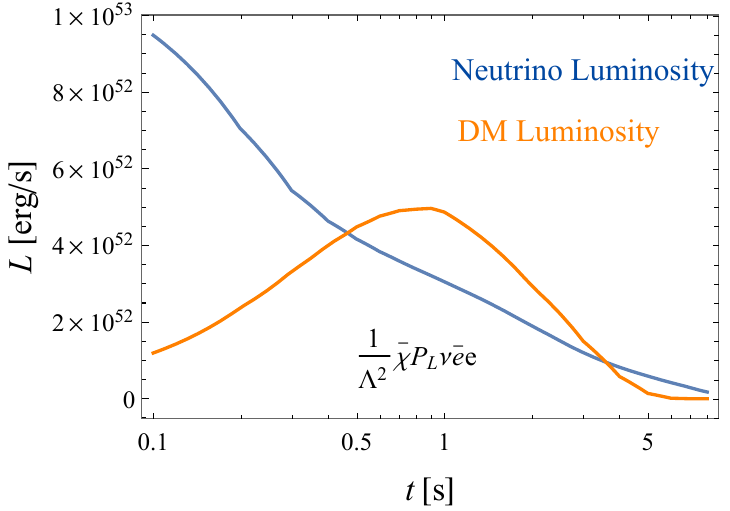}
\caption{{\it Left:} Comparison of cooling bounds obtained from the integrated energy loss criterion (gray shaded region) and the Raffelt criterion (green region) for the scalar-type operator.
{\it Right:} Time evolution of the DM luminosity (orange curve) and neutrino luminosity (blue curve).  Here we choose a scalar-type coupling to electrons with $M_\chi=1$ MeV and $\Lambda^{-1}=10^{-7}~\rm{MeV}^{-1}$ for illustration.}
\label{fig:RFandLuminosity}
\end{figure}

Here we compare two cooling criteria employed in the literature to set bounds on extra energy loss caused by new particles beyond the Standard Model. The first, often called the Raffelt criterion~\cite{Raffelt:1990yz}, requires that the luminosity of new particles at about one second after core bounce does not exceed the neutrino luminosity (i.e. $L_\chi<L_\nu \simeq 3\times10^{52}$~erg/s at 1~s). The second imposes a limit on the total energy carried away by new particles over the entire cooling phase $\mathcal{E}_\chi<0.1 \mathcal{E}_\nu$, which is argued to be more reliable than the Raffelt criterion in some of the literature~\cite{Dreiner:2013mua,Dreiner:2003wh} as it captures the full time evolution of the supernova.

Although we adopt the integrated energy loss criterion as the baseline for our analysis in this work, we perform explicit comparison of these two criteria, as illustrated in the left panel of Fig.~\ref{fig:RFandLuminosity}. Taking the scalar-type operator as an example, the shaded gray region obtained from the integrated energy loss condition is consistent with the green region from the Raffelt criterion; the Raffelt criterion yields a weaker limit by a factor of 6 compared with the integrated energy loss criterion $\mathcal{E}_\chi<0.1 \mathcal{E}_\nu$. If we choose the condition $\mathcal{E}_\chi< \mathcal{E}_\nu$ instead, the constraint will be weakened by a factor of 10 compared with the $\mathcal{E}_\chi<0.1 \mathcal{E}_\nu$, as the energy loss is proportional to the scattering cross section in the weakly-coupled regime. Similar behavior is expected for vector-type operator.

In the right panel of Fig.~\ref{fig:RFandLuminosity}, we also show the time evolution of the DM luminosity (orange curve) for $M_\chi=1$ MeV and $\Lambda^{-1}=10^{-7}~\rm{MeV}^{-1}$, and the neutrino luminosity (blue curve) extracted from the simulation data~\cite{Hudepohl:2009tyy}.  Because the DM luminosity peaks around 1~s after bounce, a luminosity comparison at that time (i.e. the Raffelet criterion) captures the critical feature of DM cooling over the period of the supernova evolution.

\section{Cooling constraints from different supernova models}
The cooling bounds derived in this work depend on the choice of the reference supernova models. To assess the robustness of cooling limits against variations in the progenitor model, we compare results using three different supernova simulations from the Garching group~\cite{Hudepohl:2009tyy,Bollig:2020xdr}:
\begin{itemize}
  \item Model A (our baseline): An 8.8 $M_\odot$ progenitor, finally leading to a 1.37 $M_\odot$ neutron star.
  \item Model B: An 18.8 $M_\odot$ progenitor (SFHo equation of state), leading to a 1.35 $M_\odot$ neutron star.
  \item Model C: A 20.0 $M_\odot$ progenitor (LS220 equation of state), leading to a 1.93 $M_\odot$ neutron star.
\end{itemize}

\begin{figure}[!htb]
\includegraphics[width=0.49\columnwidth]{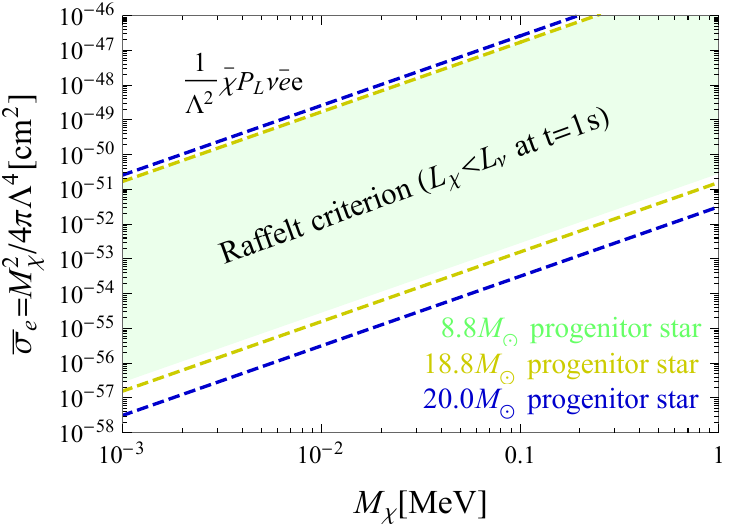}
\caption{Cooling bounds on the scalar-type interaction derived from three different supernova models: an 8.8 $M_\odot$ progenitor (green shaded region), an 18.8 $M_\odot$ progenitor (yellow dashed lines), and a 20.0 $M_\odot$ progenitor (blue dashed lines). The limits show the same overall shape and order of magnitude. The bounds from our baseline 8.8 $M_\odot$ model are the most conservative.}
\label{fig:SNmodel}
\end{figure}

Following the method described in the main text, we use the Raffelt criterion to obtain the corresponding cooling bounds for each model. In Fig.~\ref{fig:SNmodel}, the green region corresponds to model A, yellow lines to Model B, and blue lines to Model C. Despite the different models used, the upper boundaries of the exclusion regions are comparable and the lower boundaries vary by at most an order of magnitude. The limits derived from our baseline model (8.8 $M_\odot$ progenitor)~\cite{Hudepohl:2009tyy} are the most conservative, similar to the observation in~\cite{Hardy:2024gwy,Bollig:2020xdr} that larger progenitor mass leads to tighter constraints. We take scalar-type operator as an example and vector-type mediator behaves similarly.

\begin{figure}[!htb]
\includegraphics[width=0.49\columnwidth]{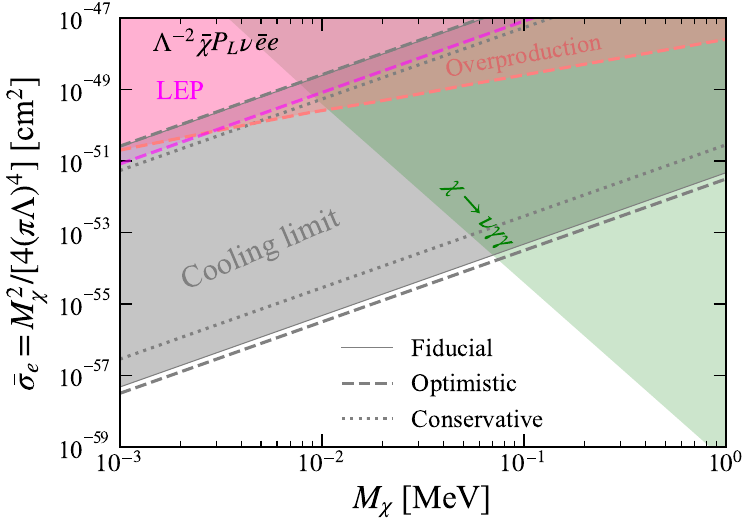}
\caption{Cooling bounds on the scalar-type interaction. Gray lines of different types correspond to the fiducial constraints in the main text, the conservative limits, and the optimistic ones. See text for details.}
\label{fig:uncertainties}
\end{figure}

To better quantify the uncertainties in the constraints, we plot three types of constraints for scalar couplings in Fig.~\ref{fig:uncertainties}: the fiducial constraints correspond to the constraint region in Fig.~\ref{fig:coolinglimit} of the main text; the optimistic and conservative ones combine the
limits from Fig.~\ref{fig:coolinglimit}. In addition, for the upper boundary of the constraint region, the conservative one takes into account the factor of 3 caused by the propagation effect. The variation of the constraints roughly reflects the uncertainties from the choices of the supernova models, cooling constraints, and the directional effects in DM propagation.

\section{The other possible cooling channels}
In addition to the $\nu e \rightarrow \chi e$ process considered in the main text, other interactions such as neutrino–neutrino scattering $\nu\nu\rightarrow\chi\chi$ or $\nu\bar{\nu}\rightarrow\chi\bar{\chi}$ could in principle also produce DM and contribute to supernova cooling from the same operators. As can be seen from Eq.~(5) in the main text, the DM production rate is proportional to the number density of the scattering target. After supernova core bounce, the electron number density is substantially higher than the neutrino number density (see Fig.~\ref{fig:simulationdata}), and the difference is a generic feature of core collapse supernovae (e.g. SFHo-18.8 and LS220-20.0). Consequently, neutrino-electron scattering $\nu e \rightarrow \chi e$ dominates DM production as long as the mediator couplings to electrons and to neutrinos are of comparable strength ($g_{\chi\nu}\approx g_{ee}$).

\begin{figure}[htbp]
\includegraphics[width=0.49\columnwidth]{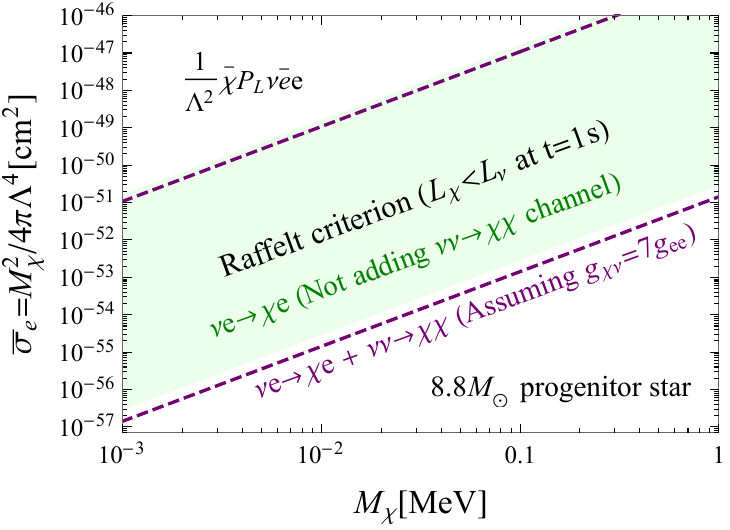}
\caption{Cooling bounds for the scalar-type interaction obtained from the Raffelt luminosity criterion. The green shaded region represents the cooling constraints only from $\nu e \rightarrow \chi e$ scattering; the purple dashed line includes $\nu\nu\rightarrow\chi\chi$ in addition with a mediator-neutrino coupling $g_{\chi \nu}$ substantially larger than the mediator-electron coupling $g_{ee}$.}
\label{fig:vvscatter}
\end{figure}

To quantify the possible impact of $\nu\nu$ scattering, we have computed the additional cooling due to $\nu\nu\rightarrow\chi\chi$ for the scalar-type operator. We have neglected DM production processes involving antineutrinos as the antineutrino number density is even smaller from Fig.~\ref{fig:simulationdata}. The calculation follows the same formalism as for the $e\nu$ case with the appropriate replacement of the target particle. Fig.~\ref{fig:vvscatter} shows the resulting cooling bounds obtained from the Raffelt luminosity criterion for two scenarios: (i) DM production only from the $e\nu$ scattering (green shaded region) and (ii) DM production from both $e\nu$ and $\nu\nu$ scattering (purple dashed line). When $g_{ee}=g_{\chi\nu}$, the contribution of $\nu\nu$ scattering is negligible and the exclusion curve coincides with the green region. Only when the neutrino coupling is taken to be much larger, i.e. $g_{\chi\nu}\approx 7g_{ee}$, the total cooling from $\nu\nu\rightarrow\chi\chi$ is roughly equal to that from $\nu e \rightarrow \chi e$, leading to a modest strengthening of the bound.

Moreover, a key motivation behind studying the cooling from $\nu e$ scattering is that, it is exactly the same operator that causes the absorption of DM in terrestrial direct detection experiments. In contrast, if $g_{\chi\nu}\gg g_{ee}$ constraints derived from neutrino scattering depend only on the neutrino–DM coupling and cannot be directly compared with direct detection experiments (unless a relationship between $g_{ee}$ and $g_{\chi\nu}$ is presumed). Our work therefore concentrates on the opposite coupling regime, which allows a clear confrontation and complementarity with present and future direct detection searches.

\section{Effects of angular dependence and dark sector interactions on dark matter propagation}
\label{sec:propagationeffects}

Here we refine our treatment of the DM propagation by evaluating two effects: (i) the angular dependence of DM trajectories and its impact on the trapping limit, and (ii) the relative importance of DM self-interactions ($\chi\chi$) and DM-neutrino interactions ($\nu\chi$) compared to DM-electron scattering ($\chi e$).

In the main text calculation of the DM survival probability against back-conversion (Eq.~(8)), we assumed for simplicity that DM particles travel on purely radial trajectories from their production point to the supernova surface. Here, we relax this assumption to estimate the effect of angular averaging.

The trapping limit is primarily determined by the optical depth of DM, which is proportional to the column density of electrons along the DM trajectory. For a DM particle produced at radius $r_0$, the column density along a radial trajectory is
\begin{equation}
I_{\rm{radial}}=\int_{r_0}^{R_{\max}} n_e(r)dr\,.
\end{equation}
We select $r_0=10$~km, where the DM luminosity is approximately maximized (see Fig.~\ref{fig:Eloss_chi}), and $R_{\max}=40$~km, beyond which the electron number density is negligible.

However, in a more realistic scenario, DM particles are emitted in all directions. Assuming isotropic emission, the angle-averaged column density is given by
\begin{equation}
\langle I\rangle=\frac{1}{4\pi} \int d\Omega \int_0^{s_{\max }(\theta)} n_e\left(\sqrt{r_0^2+s^2+2 r_0 s \cos \theta}\right) d s\ ,
\end{equation}
where $s_{\max }(\theta)=-r_0 \cos \theta+\sqrt{R_{\max}^2-r_0^2 \sin ^2 \theta}$ and $\theta$ is the angle between the initial emission direction and the outward radial vector. Evaluating this integral at $t=1$~s yields $\langle I\rangle \approx 3I_{\rm{radial}}$.

Because DM production is not strictly confined to a single radius but occurs broadly throughout the inner core ($r \lesssim 10$~km), the geometric difference between radial and angle-averaged column densities diminishes for production points closer to the center. Therefore, accounting for angular effects modifies the trapping limit by at most a factor of 3. This $\mathcal{O}(1)$ correction is negligible compared to the multiple orders of magnitude spanned by our exclusion region.

Furthermore, we consider that once DM is produced, it could potentially undergo self-scattering ($\chi\chi \rightarrow \nu\nu$) or scattering with neutrinos ($\chi\nu \rightarrow \chi\nu$), in addition to the primary $\chi e \rightarrow \nu e$ process. To assess this, we provide a quantitative estimate of the DM number density, $n_\chi$, demonstrating that these dark-sector and neutrino interactions remain subdominant to $\chi e$ scattering during DM propagation.

In the free-streaming regime, DM particles escape the supernova immediately after production. The DM number density can be estimated as the production rate multiplied by the escape time, i.e., $n_\chi \sim \dot{n}_\chi \tau_{\rm{esc}} \sim \dot{n}_\chi R_{\max}/c$. Taking representative parameters at $t=1$~s and $r=10$~km (where DM production maximizes; see Fig.~\ref{fig:Eloss_chi} and~\ref{fig:RFandLuminosity}), and using Eq.~(7) from the main text, we find $\dot{n}_\chi \sim 10^{36}~\textrm{cm}^{-3}\textrm{s}^{-1}$ at the lower boundary of the constraint region, yielding $n_\chi \sim 10^{32}~\textrm{cm}^{-3}$. This is orders of magnitude smaller than the local electron number density (see Fig.~\ref{fig:simulationdata}).

As the interaction cross section increases, the DM number density grows until it reaches full thermal equilibrium with the supernova plasma. In this limit, the density is $n_\chi=\frac{3\zeta(3)}{2\pi^2} g T^3 \sim 10^{36}~\textrm{cm}^{-3}$ for a plasma temperature of $T \sim 30$~MeV. Consequently, across the entire parameter space of the exclusion region, the DM number density never exceeds the neutrino number density, and both remain much smaller than the electron number density ($\sim 10^{38}~\textrm{cm}^{-3}$). Therefore, DM-electron scattering dominates DM propagation over $\chi\nu$ or $\chi\chi$ interaction processes, unless the mediator-neutrino (and DM) coupling $g_{\chi\nu}$ is much larger than the mediator-electron coupling $g_{ee}$ to overcome the difference in the number density.

Finally, we emphasize that the trapping limit derived in the main text is highly conservative. We assumed that a single scattering event completely removes the DM particle from the energy-loss budget, which roughly corresponds to a mean-free-path comparable to the supernova core radius. In reality, DM is only fully trapped at lower mean-free-path and the converted neutrino will subsequently scatter again to produce DM. Modeling this fully diffusive regime is beyond the scope of this work, but doing so would only serve to broaden our exclusion region, further reinforcing the robustness of our bounds.

\section{Thermal and angular averaging of DM production}
\label{sec:average}

In this section, we assess the approximate treatment of DM production in the main text by integrating over the initial electron and neutrino distributions and the outgoing scattering angles. We consider the scalar interaction and the process $e^-(p_1)+\nu(p_2)\to e^-(p_3)+\chi(p_4)$. The energy loss rate of DM per unit volume is
\begin{align}
Q
={}&
\int \prod_{i=1}^{4}
\frac{d^3\mathbf{p}_i}{(2\pi)^3\,2E_i}\,
(2\pi)^4
\delta^{(4)}(p_1+p_2-p_3-p_4)
f_e(E_1)f_\nu(E_2)
[1-f_e(E_3)]
\sum_{\rm spins}|\mathcal{M}|^2 E_4,
\label{eq:full_collision_integral}
\end{align}
The factor $E_4$ accounts for
the energy carried away in scattering. $f_i$ represents the Fermi-Dirac occupation numbers, $f_i(E)
=1/\left(\exp[(E-\mu_i)/T]+1\right)$. We simplify the
collision integral by neglecting $m_e$, $m_\nu$, and
$m_\chi$ considering $m_e$, $m_\nu$, $m_\chi \ll T$ in SN. Following Ref.~\cite{Manzari:2023gkt}, the final-state electron blocking can be replaced by its thermal average
\begin{equation}
1-f_e(E_3)\longrightarrow F_e
=
\frac{\displaystyle
\int \frac{d^3\mathbf{p}}{(2\pi)^3}
f_e(E)[1-f_e(E)]}
{\displaystyle
\int \frac{d^3\mathbf{p}}{(2\pi)^3}f_e(E)}.
\label{eq:averaged_electron_blocking}
\end{equation}
These approximations can lead to a simple analytical formula for the contributions $Q$. For isotropic distributions, integration over the overall orientation of the incoming particles gives
\begin{equation}
Q_
=
\frac{1}{32\pi^4}
\int_0^\infty dE_1\,E_1 f_e(E_1)
\int_0^\infty dE_2\,E_2 f_\nu(E_2)
\int_{-1}^{1}dc\,J_S,
\label{eq:initial_phase_space}
\end{equation}
where $c$ is the cosine of the angle between the two
incoming momenta in the supernova frame,
$s=2E_1E_2(1-c)$, and
\begin{equation}
J_S
=
\frac{F_e}{32\pi^2}
\int_{-1}^{1}dz
\int_0^{2\pi}d\phi'\,
\sum_{\rm spins}|\mathcal{M}|^2 E_4.
\label{eq:final_phase_space}
\end{equation}

The final-state phase space is evaluated in
the center-of-mass frame, however,
the DM energy $E_4$ is in the supernova frame. Let $E=E_1+E_2$, $P=|\mathbf{p}_1+\mathbf{p}_2|=\sqrt{E^2-s}$. The boost between the two frames has
$\gamma=E/\sqrt{s}$ and $\beta=P/E$.
If $\eta$ denotes the angle between the boost direction
and the incoming electron in the center-of-mass frame,
then $P\cos\eta=E_1-E_2$. Transforming the outgoing DM
energy gives
\begin{equation}
E_4
=
\frac{E}{2}
-\frac{P}{2}
\left[
\cos\eta\,z
+\sin\eta\sqrt{1-z^2}\cos\phi'
\right]\,,
\end{equation}
with the angular average
\begin{equation}
\langle E_4\rangle_{\phi'}
=\frac{E_1+E_2-(E_1-E_2)z}{2}.
\label{eq:dm_energy_angular_dependence}
\end{equation}

\begin{figure}[htbp]
\includegraphics[width=0.6\columnwidth]{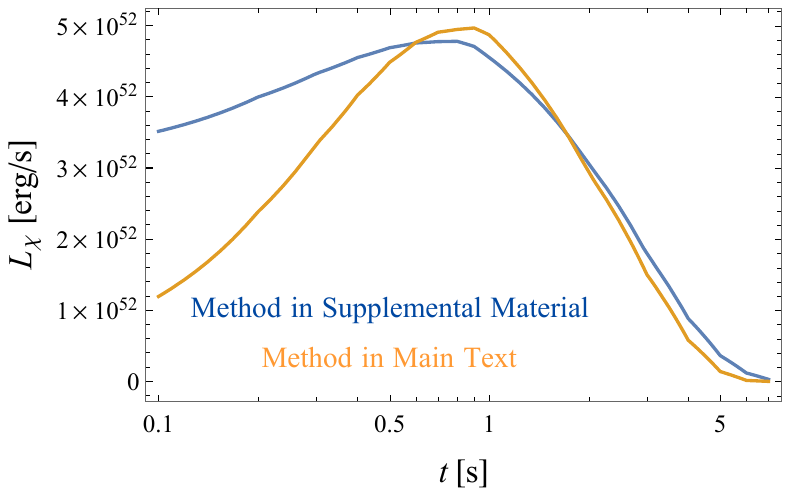}
\caption{Time evolution of the DM luminosity calculated using the two prescriptions. The orange curve
corresponds to the main text calculation, which uses
the mean electron energy and an approximate relation
between the neutrino and outgoing DM energies.
The blue curve shows the calculation presented in
this section of the Supplemental Material, which integrates over the initial state
distributions and final state phase space.
Both calculations give similar overall luminosity histories, with emission peaking around $1\,\mathrm{s}$ after core bounce and decreasing thereafter. Here we choose a scalar-type coupling to electrons with $M_\chi=1$ MeV and $\Lambda^{-1}=10^{-7}~\rm{MeV}^{-1}$ for illustration.}
\label{fig:DML}
\end{figure}

Therefore, the final-state angular integral becomes
\begin{align}
J_S
&=
\frac{F_e s^2}{32\pi\Lambda^4}
\int_{-1}^{1}dz\,
(1-z)^2
\frac{E_1+E_2-(E_1-E_2)z}{2}=
\frac{F_e s^2}{48\pi\Lambda^4}
(3E_1+E_2).
\label{eq:scalar_integrated_kernel}
\end{align}
The remaining incoming-angle integral is $\int_{-1}^{1}dc\,s^2=\frac{32}{3}E_1^2E_2^2$, which yields
\begin{equation}
Q
=
\frac{F_e}{144\pi^5\Lambda^4}
\int_0^\infty dE_1
\int_0^\infty dE_2\,
f_e(E_1)f_\nu(E_2)
\left(3E_1^4E_2^3+E_1^3E_2^4\right).
\label{eq:scalar_energy_integrals}
\end{equation}
Introducing the Fermi integrals $H_n(y)
=\int_0^\infty dx\,\frac{x^n}{1+e^{x-y}}$, $y_i=\frac{\mu_i}{T}$, we have
$\int_0^\infty dE\,E^n f_i(E)=T^{n+1}H_n(y_i)$.
Consequently, the final result can be expressed as
\begin{equation}
Q
=
\frac{T^9F_e}{144\pi^5\Lambda^4}
\left[
3H_4(y_e)H_3(y_\nu)
+
H_3(y_e)H_4(y_\nu)
\right].
\label{eq:scalar_thermal_emissivity}
\end{equation}
The total emissivity rate is summed over the neutrino
and antineutrino species $a$ and integrated over the
supernova volume,
\begin{equation}
L_\chi(t)
=\int_0^{R_{\max}}4\pi r^2dr\,
\sum_a Q_{a}(t,r).
\end{equation}

In Fig.~\ref{fig:DML}, we show the
time evolution of the DM luminosity obtained with the
main-text approximation and the energy and angular
integration described above. Both calculations give
similar overall luminosity histories, with emission
peaking around $1\,\mathrm{s}$ after core bounce and
decreasing thereafter. The time-integrated energy loss differs by 10\%, while the difference in the luminosity is 7\% at $1\,\mathrm{s}$. This indicates that the cooling bounds are only very mildly affected by
the choice of DM production prescriptions, regardless of using the integrated energy-loss criterion or the Raffelt criterion.

\section{Estimate of the nucleon Pauli blocking effect}
\label{sec:Pauliblocking}

In this section,
we estimate the size of the Pauli blocking effect of nucleons using the representative proto-neutron-star conditions of Ref.~\cite{Manzari:2023gkt}. These conditions are specified by the temperature \(T\), mass density \(\rho\), the net electron and muon number fractions per baryon \(Y_e\) and \(Y_\mu\), and the neutrino chemical potential $\mu_{\nu_e}$ respectively,
\begin{equation}
T=30~\mathrm{MeV},
\qquad
\rho=2\times10^{14}~\mathrm{g\,cm^{-3}},
\qquad
Y_e=0.12,
\qquad
Y_\mu=0.026,
\qquad
\mu_{\nu_e}=20~\mathrm{MeV}
\label{eq:typicalquantities}
\end{equation}
where we can infer the proton and neutron fractions,
$Y_p\simeq Y_e+Y_\mu\simeq0.146$,
$Y_n\simeq1-Y_p\simeq0.854$, and
$n_p\simeq0.0175~\mathrm{fm}^{-3}$,
$n_n\simeq0.102~\mathrm{fm}^{-3}$.

For each nucleon species \(N=n,p\), we employ the full finite-temperature Fermi-Dirac distributions and write
\begin{equation}
f_N(p)=
\frac{1}{
\exp\left[
\left(E_N^{\rm kin}(p)-\mu_N^{\rm kin}\right)/T
\right]+1},
\label{eq:FD_nucleon}
\end{equation}
where
$E_N^{\rm kin}(p)=\sqrt{p^2+m_N^2}-m_N$,
$\mu_N^{\rm kin}=\mu_N^{\rm tot}-m_N$ .
The kinetic chemical potential is determined by requiring the Fermi-Dirac distribution to reproduce the corresponding nucleon number density,
\begin{equation}
n_N= 2\int\frac{d^3p}{(2\pi)^3}\,f_N(p)\,.
\label{eq:nucleon_density_FD}
\end{equation}
Solving Eq.~(\ref{eq:nucleon_density_FD}) numerically at \(T=30~\mathrm{MeV}\) for the densities above gives
\begin{equation}
\mu_n^{\rm kin}\simeq18.7~\mathrm{MeV},
\qquad
\mu_p^{\rm kin}\simeq-44.4~\mathrm{MeV}.
\end{equation}
The corresponding degeneracy parameters are
\begin{equation}
\eta_n\equiv\frac{\mu_n^{\rm kin}}{T}\simeq0.62,
\qquad
\eta_p\equiv\frac{\mu_p^{\rm kin}}{T}\simeq-1.48.
\end{equation}
Thus the neutrons are only mildly to moderately degenerate, while the protons are nondegenerate. This is also reflected in the zero-momentum occupations,
$f_n(0)\simeq0.65$,
and
$f_p(0)\simeq0.19$.

We estimate the final-state Pauli suppression for momentum transfer \({\bf q}\) through
\begin{equation}
B_N(q,T)
=
\frac{
\displaystyle
\int d^3p\,f_N(p)
\left[
1-f_N(|{\bf p}+{\bf q}|)
\right]
}{
\displaystyle
\int d^3p\,f_N(p)
}.
\label{eq:pauli_B}
\end{equation}
The factor \(1-f_N\) is the probability that the final nucleon state is unoccupied, while \(B_N=1\) corresponds to the absence of Pauli blocking.
To estimate a characteristic momentum transfer, we use the thermal \(\nu_e\) distribution with the chemical potential in Eq.~\eqref{eq:typicalquantities} and find the average neutrino energy
$\langle E_{\nu_e}\rangle\simeq98~\mathrm{MeV}$.
For \(m_\chi\ll E_\nu\) and small nucleon recoil,
$q\simeq
2E_\nu\sin\frac{\theta}{2}$.
For an approximately isotropic contact interaction, angular averaging gives
$\langle q\rangle
=
\frac{4}{3}\langle E_{\nu_e}\rangle
\simeq130~\mathrm{MeV}$.
Evaluating Eq.~(\ref{eq:pauli_B}) with the finite-temperature distributions in Eq.~(\ref{eq:FD_nucleon}) at \(q=130~\mathrm{MeV}\) gives
$B_n\simeq0.74$,
$B_p\simeq0.94$.
For comparable couplings to protons and neutrons, a density-weighted estimate gives
$B_{\rm eff}=
Y_nB_n+Y_pB_p
\simeq0.77$.
The total nucleon production rate is therefore reduced by approximately \(23\%\) under these representative PNS conditions.

%%%%%%%%%%%%%%%%%%%%%%%%%%%%%%%%%%%%%%%%%%%%%%%%%%%%%%%%
\end{document}